\documentclass[apj, twocolumn, twocolappendix, numberedappendix, appendixfloats]{openjournal}

\usepackage[T1]{fontenc}
\usepackage[utf8]{inputenc}

\usepackage{graphicx}
\graphicspath{{./}{Figures/}{Plots/}}

\usepackage[pdfpagelabels=false]{hyperref}

\usepackage[dvipsnames]{xcolor}
\usepackage{xspace}
\usepackage{soul}
\usepackage{booktabs}
\usepackage{adjustbox}
\usepackage{needspace}

\usepackage{amsmath}
\usepackage{amssymb}
\usepackage{commath}
\usepackage{upgreek}

\usepackage{savesym}
\savesymbol{tablenum}
\usepackage{siunitx}
\restoresymbol{SIX}{tablenum}

\usepackage{nth}

\usepackage[capitalise, noabbrev]{cleveref}
\usepackage{enumitem}
\usepackage{csvsimple}
\usepackage{multirow}
\usepackage{array}
\usepackage{placeins}

\crefname{figure}{Fig.}{Figs.}
\crefname{equation}{equation}{equations}
\crefname{appendix}{Appendix}{Appendices}

\bibpunct[]{(}{)}{;}{a}{}{,}

\DeclareRobustCommand{\VAN}[3]{#2}
\let\VANthebibliography\thebibliography
\def\thebibliography{\DeclareRobustCommand{\VAN}[3]{##3}\VANthebibliography}

\newcommand{\orcidsymb}[2]{#1\href{http://orcid.org/#2}{\adjustbox{trim={-.15\width} {0\height} {-.15\width} {0\height},clip}{\includegraphics[height=10pt]{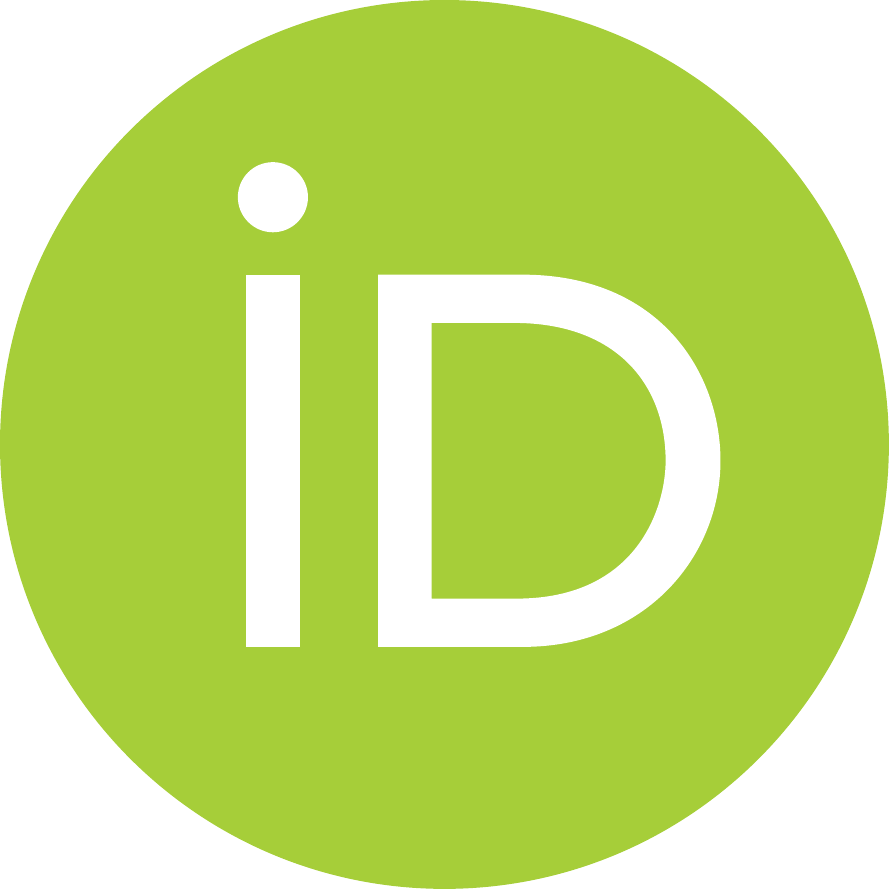}}}}
\newcommand{\orcidauthor}[3]{\author{\orcidsymb{#1}{#2}$^{#3}$}}

\newcommand{\Angstrom}{\text{\AA}}
\newcommand{\ssim}{\sim \!}

\newcommand{\Lymana}{{Lyman-\ensuremath{\upalpha}}\xspace}

\newcommand{\Lya}{{Ly\ensuremath{\upalpha}}\xspace}

\newcommand{\HI}{\hbox{H\,{\sc i}}\xspace}
\newcommand{\HII}{\hbox{H\,{\sc ii}}\xspace}
\newcommand{\CIV}{\hbox{C\,{\sc iv}}\xspace}

\newcommand{\CII}{\hbox{[C\,{\sc ii}]}\xspace}

\newcommand{\OII}{\hbox{[O\,{\sc ii}]}\xspace}
\newcommand{\OIII}{\hbox{[O\,{\sc iii}]}\xspace}

\newcommand{\NeIII}{\hbox{[Ne\,{\sc iii}]}\xspace}

\newcommand{\CIIonefiftyeight}{{\ensuremath{\CII \, 158 \, \mathrm{\upmu m}}}\xspace}
\newcommand{\OIIIeightyeight}{{\ensuremath{\OIII \, 88 \, \mathrm{\upmu m}}}\xspace}

\newcommand{\Halpha}{\ensuremath{\mathrm{H}\upalpha}\xspace}
\newcommand{\Hbeta}{\ensuremath{\mathrm{H}\upbeta}\xspace}

\newcommand{\JGSzeleven}{JADES-GS-z11-0\xspace}

\hypersetup{
    unicode=true,
    final=true,
    plainpages=false,
    pdfstartview=FitV,
    pdftoolbar=false,
    pdfmenubar=true,
    bookmarksopen=true,
    bookmarksnumbered=true,
    breaklinks=true,
    linktocpage,
    colorlinks=true,
    linkcolor=Blue,
    urlcolor=Blue,
    citecolor=Blue,
    anchorcolor=green
}
\AtBeginDocument{
    \hypersetup{
        pdftitle={On the origins of oxygen},
        pdfauthor={J. Witstok et al.},
        pdfsubject={ALMA and JWST characterise the multi-phase, metal-enriched, star-bursting medium within a 'normal' z > 11 galaxy}
    }
}

\usepackage{etoolbox}
\makeatletter
\pretocmd\@sect{\def\@currentcounter{#1}}{}{\fail}
\patchcmd\H@refstepcounter{\protected@edef}{\protected@xdef}{}{}
\makeatother

\shorttitle{ALMA and JWST characterise the multi-phase medium within a `normal' $z > 11$ galaxy}
\shortauthors{J. Witstok et al.}

\begin{document}
\title{On the origins of oxygen: ALMA and JWST characterise the multi-phase, metal-enriched, star-bursting medium within a `normal' $\MakeLowercase{z} > 11$ galaxy\vspace{-9ex}}

\orcidauthor{Joris~Witstok}{0000-0002-7595-121X}{\hyperlink{inst:DAWN}{1}, \hyperlink{inst:NBI}{2}, *}
\orcidauthor{Renske~Smit}{0000-0001-8034-7802}{\hyperlink{inst:LJMU}{3}}
\orcidauthor{William~M.~Baker}{0000-0003-0215-1104}{\hyperlink{inst:DARK}{4}}
\orcidauthor{Pierluigi~Rinaldi}{0000-0002-5104-8245}{\hyperlink{inst:Steward}{5}}
\orcidauthor{Kevin~N.~Hainline}{0000-0003-4565-8239}{\hyperlink{inst:Steward}{5}}
\orcidauthor{Hiddo~S.~B.~Algera}{0000-0002-4205-9567}{\hyperlink{inst:ASIAA}{6}}
\orcidauthor{Santiago~Arribas}{0000-0001-7997-1640}{\hyperlink{inst:CAB}{7}}
\orcidauthor{Tom~J.~L.~C.~Bakx}{0000-0002-5268-2221}{\hyperlink{inst:Chalmers}{8}, \hyperlink{inst:Nagoya}{9}, \hyperlink{inst:NAOJ}{10}}
\orcidauthor{Andrew~J.~Bunker}{0000-0002-8651-9879}{\hyperlink{inst:Oxford}{11}}
\orcidauthor{Stefano~Carniani}{0000-0002-6719-380X}{\hyperlink{inst:SNS}{12}}
\orcidauthor{Stéphane~Charlot}{0000-0003-3458-2275}{\hyperlink{inst:IAP}{13}}
\orcidauthor{Jacopo~Chevallard}{0000-0002-7636-0534}{\hyperlink{inst:Oxford}{11}}
\orcidauthor{Mirko~Curti}{0000-0002-2678-2560}{\hyperlink{inst:ESO}{14}}
\orcidauthor{Emma~Curtis-Lake}{0000-0002-9551-0534}{\hyperlink{inst:Herts}{15}}
\orcidauthor{Daniel~J.~Eisenstein}{0000-0002-2929-3121}{\hyperlink{inst:CfA}{16}}
\orcidauthor{Kasper~E.~Heintz}{0000-0002-9389-7413}{\hyperlink{inst:DAWN}{1}, \hyperlink{inst:NBI}{2}, \hyperlink{inst:Geneva}{17}}
\orcidauthor{Jakob~M.~Helton}{0000-0003-4337-6211}{\hyperlink{inst:Steward}{5}}
\orcidauthor{Gareth~C.~Jones}{0000-0002-0267-9024}{\hyperlink{inst:Kavli}{18}, \hyperlink{inst:Cavendish}{19}}
\orcidauthor{Roberto~Maiolino}{0000-0002-4985-3819}{\hyperlink{inst:Kavli}{18}, \hyperlink{inst:Cavendish}{19}, \hyperlink{inst:UCL}{20}}
\orcidauthor{Michael~V.~Maseda}{0000-0003-0695-4414}{\hyperlink{inst:Wisconsin}{21}}
\orcidauthor{Pablo~G.~Pérez-González}{0000-0003-4528-5639}{\hyperlink{inst:CAB}{7}}
\orcidauthor{Clara~L.~Pollock}{0009-0001-2808-4918}{\hyperlink{inst:DAWN}{1}, \hyperlink{inst:NBI}{2}}
\orcidauthor{Brant~E.~Robertson}{0000-0002-4271-0364}{\hyperlink{inst:UCSC}{22}}
\orcidauthor{Aayush~Saxena}{0000-0001-5333-9970}{\hyperlink{inst:Oxford}{11}, \hyperlink{inst:UCL}{20}}
\orcidauthor{Jan~Scholtz}{0000-0001-6010-6809}{\hyperlink{inst:Kavli}{18}, \hyperlink{inst:Cavendish}{19}}
\orcidauthor{Irene~Shivaei}{0000-0003-4702-7561}{\hyperlink{inst:CAB}{7}}
\orcidauthor{Fengwu~Sun}{0000-0002-4622-6617}{\hyperlink{inst:CfA}{16}}
\orcidauthor{Sandro~Tacchella}{0000-0002-8224-4505}{\hyperlink{inst:Kavli}{18}, \hyperlink{inst:Cavendish}{19}}
\orcidauthor{Hannah~Übler}{0000-0003-4891-0794}{\hyperlink{inst:MPE}{23}}
\orcidauthor{Darach~Watson}{0000-0002-4465-8264}{\hyperlink{inst:DAWN}{1}, \hyperlink{inst:NBI}{2}}
\orcidauthor{Chris~J.~Willott}{0000-0002-4201-7367}{\hyperlink{inst:NRC}{24}}
\orcidauthor{Zihao~Wu}{0000-0002-8876-5248}{\hyperlink{inst:CfA}{16}}

\affiliation{\hypertarget{inst:DAWN}$^{1}$Cosmic Dawn Center (DAWN), Copenhagen, Denmark}
\affiliation{\hypertarget{inst:NBI}$^{2}$Niels Bohr Institute, University of Copenhagen, Jagtvej 128, DK-2200, Copenhagen, Denmark}
\affiliation{\hypertarget{inst:LJMU}$^{3}$Astrophysics Research Institute, Liverpool John Moores University, 146 Brownlow Hill, Liverpool L3 5RF, UK}
\affiliation{\hypertarget{inst:DARK}$^{4}$DARK, Niels Bohr Institute, University of Copenhagen, Jagtvej 155A, DK-2200 Copenhagen, Denmark}
\affiliation{\hypertarget{inst:Steward}$^{5}$Steward Observatory, University of Arizona, 933 N. Cherry Avenue, Tucson AZ 85721, USA}
\affiliation{\hypertarget{inst:ASIAA}$^{6}$Institute of Astronomy and Astrophysics, Academia Sinica, 11F of Astronomy-Mathematics Building, No. 1, Sec. 4, Roosevelt Rd, Taipei 106216, Taiwan, ROC}
\affiliation{\hypertarget{inst:CAB}$^{7}$Centro de Astrobiolog\'ia (CAB), CSIC–INTA, Cra. de Ajalvir Km.~4, 28850- Torrej\'on de Ardoz, Madrid, Spain}
\affiliation{\hypertarget{inst:Chalmers}$^{8}$Department of Space, Earth, \& Environment, Chalmers University of Technology, Chalmersplatsen, SE-4 412 96 Gothenburg, Sweden}
\affiliation{\hypertarget{inst:Nagoya}$^{9}$Department of Physics, Graduate School of Science, Nagoya University, Aichi 464-8602, Japan}
\affiliation{\hypertarget{inst:NAOJ}$^{10}$National Astronomical Observatory of Japan, 2-21-1, Osawa, Mitaka, Tokyo 181-8588, Japan}
\affiliation{\hypertarget{inst:Oxford}$^{11}$Department of Physics, University of Oxford, Denys Wilkinson Building, Keble Road, Oxford OX1 3RH, UK}
\affiliation{\hypertarget{inst:SNS}$^{12}$Scuola Normale Superiore, Piazza dei Cavalieri 7, I-56126 Pisa, Italy}
\affiliation{\hypertarget{inst:IAP}$^{13}$Sorbonne Universit\'e, CNRS, UMR 7095, Institut d'Astrophysique de Paris, 98 bis bd Arago, 75014 Paris, France}
\affiliation{\hypertarget{inst:ESO}$^{14}$European Southern Observatory, Karl-Schwarzschild-Strasse 2, 85748 Garching, Germany}
\affiliation{\hypertarget{inst:Herts}$^{15}$Centre for Astrophysics Research, Department of Physics, Astronomy and Mathematics, University of Hertfordshire, Hatfield AL10 9AB, UK}
\affiliation{\hypertarget{inst:CfA}$^{16}$Center for Astrophysics $|$ Harvard \& Smithsonian, 60 Garden St., Cambridge MA 02138, USA}
\affiliation{\hypertarget{inst:Geneva}$^{17}$Department of Astronomy, University of Geneva, Chemin Pegasi 51, 1290 Versoix, Switzerland}
\affiliation{\hypertarget{inst:Kavli}$^{18}$Kavli Institute for Cosmology, University of Cambridge, Madingley Road, Cambridge CB3 0HA, UK}
\affiliation{\hypertarget{inst:Cavendish}$^{19}$Cavendish Laboratory, University of Cambridge, 19 JJ Thomson Avenue, Cambridge CB3 0HE, UK}
\affiliation{\hypertarget{inst:UCL}$^{20}$Department of Physics and Astronomy, University College London, Gower Street, London WC1E 6BT, UK}
\affiliation{\hypertarget{inst:Wisconsin}$^{21}$Department of Astronomy, University of Wisconsin-Madison, 475 N. Charter St., Madison WI 53706, USA}
\affiliation{\hypertarget{inst:UCSC}$^{22}$Department of Astronomy and Astrophysics University of California, Santa Cruz, 1156 High Street, Santa Cruz CA 96054, USA}
\affiliation{\hypertarget{inst:MPE}$^{23}$Max-Planck-Institut für extraterrestrische Physik (MPE), Gießenbachstraße 1, 85748 Garching, Germany}
\affiliation{\hypertarget{inst:NRC}$^{24}$NRC Herzberg, 5071 West Saanich Rd, Victoria BC V9E 2E7, Canada}

\thanks{$^*$E-mail: \href{mailto:joris.witstok@nbi.ku.dk}{joris.witstok@nbi.ku.dk}}

\begin{abstract}
    The unexpectedly high abundance of galaxies at $z > 11$ revealed by JWST has sparked a debate on the nature of early galaxies and the physical mechanisms regulating their formation. The Atacama Large Millimeter/submillimeter Array (ALMA) has begun to provide vital insights on their gas and dust content, but so far only for extreme `blue monsters'. Here we present new, deep ALMA observations of \JGSzeleven, a more typical (sub-$L^*$) $z > 11$ galaxy that bridges the discovery space of JWST and the Hubble Space Telescope. These data confirm the presence of the \OIIIeightyeight line at $4.5\sigma$ significance, precisely at the redshift of several faint emission lines previously seen with JWST/NIRSpec, while the underlying dust continuum remains undetected ($F_\nu < 9.0 \, \mathrm{\upmu Jy}$), implying an obscured star formation rate (SFR) of $\text{SFR}_\text{IR} \lesssim 6 \, \mathrm{M_\odot \, yr^{-1}}$ and dust mass of $M_\text{dust} \lesssim 1.0 \times 10^{6} \, \mathrm{M_\odot}$ (all $3\sigma$). The accurate ALMA redshift of $z_\text{\OIII} = 11.1221 \pm 0.0006$ ($\gtrsim \! 5\times$ refined over NIRSpec) helps confirm that redshifts measured purely from the \Lymana break, even spectroscopically, should properly take into account the effects of potential damped \Lymana (DLA) absorption to avoid systematic overestimates of up to $\Delta z \approx 0.5$. The \OIIIeightyeight luminosity of $L_\text{\OIII} = (1.1 \pm 0.3) \times 10^{8} \, \mathrm{L_\odot}$, meanwhile, agrees well with the scaling relation for local metal-poor dwarfs given the SFR measured by NIRCam, NIRSpec, and MIRI. The spatially resolved MIRI and ALMA emission also underscores that \JGSzeleven is likely to consist of two low-mass components that are undergoing strong bursts of star formation yet are already pre-enriched in oxygen ($\ssim 20$-$30\%$ solar), only $400 \, \mathrm{Myr}$ after the Big Bang.\vspace{5ex}
\end{abstract}

\needspace{2.5cm}
\section{Introduction}
\label{sec:Introduction}

Across several decades, the Hubble Space Telescope (HST) has made remarkable strides in finding evermore distant galaxies \citep[for a review, see][]{2016ARA&A..54..761S}. It was quickly realised that the suppression of ultraviolet (UV) light due to residual neutral, atomic hydrogen (\HI) in the intervening intergalactic medium (IGM) could be exploited to efficiently identify sources at redshifts $z \gtrsim 2$ \citep{1996ApJ...462L..17S, 1996MNRAS.283.1388M}. The distinguishing `Lyman break' universally observed in the spectral energy distribution (SED) of these sources becomes particularly pronounced entering the epoch of reionisation (EoR) at redshifts $z > 6$ \citep{2004MNRAS.355..374B, 2010MNRAS.409..855B, 2015ApJ...803...34B, 2015ApJ...810...71F}, where IGM absorption fully saturates up to the rest-frame wavelength of \HI\ \Lymana (\Lya), $\lambda_\text{\Lya} = 1215.67 \, \Angstrom$ \citep{2014MNRAS.442.1805I}. However, a fundamental frontier remained at $z \approx 11$, where the \Lya transition becomes redshifted beyond the reach of HST.

Furthermore, ground-based spectrographs struggled to validate redshifts of $z \gtrsim 6$ \citep[only occasionally detecting the \Lya line;][]{2014ApJ...795...20S, 2015ApJ...810L..12Z, 2018A&A...619A.147P}, and the spectroscopic capability of HST in this redshift regime was limited. Instead, this role was partially fulfilled by the Atacama Large Millimeter/submillimeter Array (ALMA), which is able to simultaneously probe the ionisation state, metal and dust content, and kinematics of the interstellar medium (ISM) in reionisation-era galaxies \citep[see][ for a review]{2020RSOS....700556H}. Far-infrared (FIR) dust-continuum emission further indicated that these sources may rapidly build up considerable dust reservoirs \citep[e.g.][]{2015Natur.519..327W, 2022MNRAS.512..989D, 2022MNRAS.515.1751W, 2023MNRAS.523.3119W, 2023MNRAS.518.6142A, 2024MNRAS.527.6867A, 2025arXiv250606418S}. Though the \CIIonefiftyeight fine-structure line proved a versatile tracer of ionised and neutral gas \citep{2018Natur.553..178S, 2020A&A...643A...1L, 2021ApJ...922..147H, 2022ApJ...931..160B, 2022ApJ...928...31S, 2023ApJ...954..103S}, its luminosity was consistently matched or even outshone by the \OIIIeightyeight transition at $z \gtrsim 6$ \citep{2016Sci...352.1559I, 2018Natur.557..392H, 2019ApJ...874...27T}. This suggested the presence of a diffuse, oxygen-enriched ISM excited into a highly ionized state by intense radiation fields \citep{2020ApJ...896...93H, 2022MNRAS.515.1751W}, and hinted at a decreased C/O abundance ratio and potentially top-heavy initial mass function \citep[IMF;][]{2022MNRAS.510.5603K}.

Breaking the HST redshift barrier finally became possible with JWST \citep{2023PASP..135e8001M, 2023PASP..135d8001R}, which is capable of near-infrared imaging and spectroscopy at unprecedented spatial resolution and sensitivity \citep{2022ARA&A..60..121R, 2023PASP..135f8001G}. Based on Near-Infrared Camera \citep[NIRCam;][]{2023PASP..135b8001R} imaging, a number of $z \gtrsim 11$ galaxy candidates were identified soon after the launch of JWST \citep[e.g.][]{2022ApJ...938L..15C, 2022ApJ...940L..55F, 2023ApJ...946L..13F, 2022ApJ...940L..14N, 2023MNRAS.518.6011D, 2023MNRAS.518.4755A, 2023ApJS..265....5H, 2023ApJ...951L...1P, 2024ApJ...964...71H}. Indeed, the sheer number of UV-bright galaxy candidates beyond $z = 11$ hugely eclipsed $\Lambda$CDM-anchored forecasts calibrated on HST observations \citep{2023MNRAS.521..497M}, raising questions about the possible contribution of active galactic nuclei \citep[AGN;][]{2024JCAP...08..025H}, the mode of star formation becoming more bursty \citep[e.g.][]{2023MNRAS.525.3254S, 2023MNRAS.526.2665S}, less obscured by dust \citep[e.g.][]{2023MNRAS.522.3986F} and/or simply more efficient \citep[e.g.][]{2024ApJ...975..192G}.

Efforts to follow up JWST-selected candidates at the redshift frontier with ALMA, however, did not immediately yield unambiguous confirmations \citep{2023ApJ...955..130F, 2023A&A...671A..29K, 2023ApJ...950...61Y, 2023MNRAS.519.5076B, 2023A&A...669L...8P, 2023MNRAS.520L..16K}, initially calling into question the robustness of our selection of high-redshift candidates, especially those at the bright end \citep[e.g.][]{2023ApJ...943L...9Z}. The Near-Infrared Spectrograph \citep[NIRSpec;][]{2022A&A...661A..80J, 2022A&A...661A..81F} did finally deliver spectroscopic confirmations, though in many cases solely based on the \Lya spectral break \citep{2023NatAs...7..622C, 2023Natur.622..707A, 2023ApJ...957L..34W, 2024Natur.633..318C}.

In several cases, the presence of metals was revealed through UV cooling lines \citep[e.g.][]{2023A&A...677A..88B, 2024ApJ...973....8H, 2024A&A...689A.152D, 2024ApJ...972..143C, 2025A&A...693A..50N, 2025arXiv250708245T, 2025arXiv250511263N}, from which it also became apparent that redshifts based on the \Lya spectral break---even spectroscopically---were systematically overestimated \citep[e.g.][]{2023ApJ...949L..25F, 2025ApJ...983L...2A}. \Lya damping-wing absorption arising in a fully neutral IGM \citep[e.g.][]{2024MNRAS.532.1646K, 2024MNRAS.531L..34K, 2025A&A...697A..89C, 2025arXiv250111702M, 2025ApJ...987...82H} will certainly produce a smoother \Lya break, as will an increasingly prominent two-photon contribution to nebular-continuum emission expected for metal-poor stellar populations \citep{2024MNRAS.534..523C, 2025OJAp....8E.104K}.

Still, detailed modelling of the break in NIRSpec/PRISM spectra of $z \gtrsim 9$ galaxies frequently requires damped \Lya (DLA) absorption far exceeding the IGM damping-wing absorption \citep{2024ApJ...976..160H, 2025Natur.639..897W, 2024Sci...384..890H, 2025ApJ...987L...2H, 2025A&A...693A..60H}. This suggests that our sightlines may occasionally pierce substantial \HI gas reservoirs within the ISM or circumgalactic medium, as is indeed expected from simulations \citep{2025arXiv251001315G}.

Recent results indicate that the overabundance of $z \gtrsim 11$ galaxies has stood the test of time \citep[e.g.][]{2024ApJ...960...56H}, and even extends to the faint end of the UV luminosity function \citep[e.g.][]{2023ApJ...951L...1P, 2023ApJ...954L..46L, 2025ApJ...992...63W}. Meanwhile, breakthrough ALMA confirmations at the redshift frontier finally arrived by means of the \OIIIeightyeight transition \citep{2024ApJ...977L...9Z, 2025ApJ...988...19S, 2025A&A...696A..87C}, offering unique new insights into the nature of the earliest galaxies. However, so far these were restricted to the very UV-brightest galaxies known---`blue monsters'---some of which have been shown to exhibit signatures indicative of an exceptionally compact mode of star formation and/or an AGN \citep[e.g.][]{2023MNRAS.523.3516C, 2024Natur.627...59M, 2024ApJ...972..143C, 2025A&A...695A.250A}, while it is unclear whether this is the case for the fainter galaxies, which are significantly more numerous.

In this work, we present newly obtained ALMA observations of \OIIIeightyeight and dust-continuum emission in a more typical, sub-$L^*$ galaxy at $z \approx 11$. Originally discovered as UDFj-39546284 \citep{2011Natur.469..504B} and assigned a redshift of $z_\text{phot} = 11.9_{-0.5}^{+0.3}$ \citep{2013ApJ...763L...7E}, it was validated as a promising candidate in NIRCam imaging obtained as part of the JWST Advanced Deep Extragalactic Survey \citep[JADES;][]{2020IAUS..352..337R, 2020IAUS..352..342B, 2023arXiv230602465E} with $z_\text{phot} = 11.7_{-0.4}^{+0.5}$ \citep{2023NatAs...7..611R}. It subsequently became known as \JGSzeleven, one of the four first spectroscopically confirmed $z > 10$ galaxies reported by \citet{2023NatAs...7..622C}, with a redshift of $z_\text{spec, IGM} = 11.48_{-0.08}^{+0.03}$ taking into account IGM damping-wing absorption. Finally, \citet{2024ApJ...976..160H} verified its redshift based on the detection of multiple, faint UV and optical emission lines---$\CIV \, \lambda \, 1548, 1551 \, \Angstrom$ (\CIV hereafter), $\OII \, \lambda \, 3727, 3730 \, \Angstrom$ (\OII), and $\NeIII \, \lambda \, 3870 \, \Angstrom$ (\NeIII), combined at $94\%$ confidence---in ultra-deep JWST/NIRSpec observations comprising $75 \, \mathrm{h}$ in PRISM mode. Surprisingly, this placed its redshift at a refined value\footnote{As the fainter, southern counterpart to GN-z11 \citep{2016ApJ...819..129O}, \JGSzeleven is thus likely the most distant galaxy discovered by HST, with GN-z11 now confirmed at $z_\text{spec} = 10.6$ \citep{2023A&A...677A..88B}.} of $z_\text{spec} = 11.122_{-0.003}^{+0.005}$, necessitating DLA absorption in addition to the IGM damping-wing absorption, with an \HI column density of $N_\text{\HI} \approx 10^{22.5} \, \mathrm{cm^{-2}}$. Most recently, in deep JWST imaging with the Mid-Infrared Instrument \citep[MIRI;][]{2015PASP..127..584R}, \JGSzeleven was seen to exhibit a strong excess at $5.6 \, \mathrm{\upmu m}$ possibly dominated by bright $\OIII \, \lambda \, 4960, 5008 \, \Angstrom$ and $\Hbeta$ ($\OIII + \Hbeta$) emission lines \citep{2025A&A...696A..57O}.

The outline of this work is as follows. In \cref{sec:Observations}, we discuss the observations underlying this work. Our results are presented in \cref{sec:Results} and discussed in \cref{sec:Discussion}, with \cref{sec:Conclusions} providing a summary. We adopt a flat $\Lambda$CDM cosmology throughout, with $H_0 = 67.4 \, \mathrm{km \, s^{-1} \, Mpc^{-1}}$, $\Omega_\text{m} = 0.315$, $\Omega_\text{b} = 0.0492$ based on the latest results of the Planck collaboration \citep{2020A&A...641A...6P}. On-sky separations of $1\arcsec$ and $1\arcmin$ at $z = 11.1$ correspond to $3.95 \, \text{physical kpc}$ (pkpc) and $0.237 \, \text{physical Mpc}$ (pMpc), respectively. Quoted magnitudes are in the AB system \citep{1983ApJ...266..713O}, emission-line wavelengths in vacuum, and FIR luminosities are expressed in terms of the bolometric solar luminosity, $L_\odot = 3.828 \times 10^{33} \, \mathrm{erg \, s^{-1}}$.

\needspace{2.5cm}
\section{Observations}
\label{sec:Observations}

\subsection{HST and JWST observations}
\label{ssec:Observations:HST_JWST}

The NIRCam imaging and NIRSpec spectroscopy of \JGSzeleven are associated with JWST guaranteed time observations (GTO) programme IDs (PIDs) 1180 (PI: Eisenstein) and 1210 (PI: Luetzgendorf), further complemented by several JWST general observer (GO) programmes in the Hubble Ultra Deep Field \citep[HUDF;][]{2006AJ....132.1729B}. Specifically, we made use of the JWST Extragalactic Medium-band Survey \citep[JEMS;][]{2023ApJS..268...64W}, associated with PID 1963 (PIs: Williams, Maseda \& Tacchella); the First Reionization Epoch Spectroscopic COmplete Survey \citep[FRESCO;][]{2023MNRAS.525.2864O}, associated with PID 1895 (PI: Oesch); and the JADES Origins Field \citep[JOF;][]{2025ApJS..281...50E}, associated with PID 3215 (PIs: Eisenstein \& Maiolino). Compilations of these data\footnote{Publicly available at \url{https://archive.stsci.edu/hlsp/jades}, where \JGSzeleven appears under the NIRSpec IDs in \cref{tab:Source_properties}.} were previously presented in \citet{2023NatAs...7..622C}, \citet{2023NatAs...7..611R}, and \citet{2024ApJ...976..160H}. The JADES survey strategy and data reduction procedures are described in the survey overview \citep{2023arXiv230602465E} and data release papers \citep{2023ApJS..269...16R, 2024A&A...690A.288B, 2025ApJS..277....4D, 2025arXiv251001033C, 2025arXiv251001034S}. We also considered publicly available HST imaging \citep{2016arXiv160600841I} over the Great Observatories Origins Deep Survey-South \citep[GOODS-S;][]{2004ApJ...600L..93G} legacy field. Finally, we made use of the publicly available MIRI imaging\footnote{Publicly available at \url{https://doi.org/10.5281/zenodo.15535601} \citep{Melinder2025}. Shallower coverage in all MIRI filters was taken as part of the Systematic Mid-infrared Instrument Legacy Extragalactic Survey \citep[SMILES;][]{2024ApJ...976..224A}, though \JGSzeleven remained undetected.} in the F560W, F770W, and F1000W filters from the MIRI Deep Imaging Survey \citep[MIDIS;][, in prep.]{2025A&A...696A..57O}, associated with PID 1283 (PI: \"Ostlin).

The NIRCam imaging presented here was reduced with the latest v1.0 data reduction pipeline. Compared to previous versions \citep[e.g. v0.9 in][]{2024ApJ...976..160H}, improvements at the low-level processing in this version include wisp amelioration, astrometry, and noise mitigation, whereas high-level improvements include image weighting during mosaicing, overall depth and area of the imaging, as will be described in an upcoming paper \citetext{B.~D.~Johnson \& JADES Collaboration in prep.}.
\begin{figure*}
	\centering
	\includegraphics[width=\linewidth]{"NIRCam_NIRSpec"}
	\caption{NIRCam and NIRSpec/PRISM observations of \JGSzeleven. \textbf{a}, An inverse-variance weighted stack of PSF-matched NIRCam images from all filters with firm continuum detections, starting at F182M and going redwards (see Appendix~\ref{app:Photometric_measurements} for details). Black and white crosses show the NIRCam-based centroids of the two components (A and B) of \JGSzeleven. The placement of the NIRSpec micro-shutters is shown in white (nearly identical across all visits). A horizontal bar indicates a physical scale of $1 \, \mathrm{kpc}$ at $z = 11.122$. \textbf{b}, One-dimensional NIRSpec/PRISM spectrum. Shading shows $1 \sigma$ uncertainty on individual wavelength bins (i.e. the covariance matrix diagonal, $\sigma_i^2 = C_{ii}$; see Appendix~\ref{app:Covariance_matrix}). For the redshift solution $z = 11.122$ reported by \citet{2024ApJ...976..160H}, the location of key rest-frame UV and optical emission lines is indicated, most prominent among which are the \CIV, \OII, and \NeIII emission lines.
	}
	\label{fig:NIRCam_NIRSpec}
\end{figure*}
\begingroup
    \setlength{\tabcolsep}{6pt} 
    \renewcommand{\arraystretch}{1.25} 
    \begin{deluxetable}{lcc}
        \tabletypesize{\footnotesize}
        \tablecaption{General properties of \JGSzeleven\label{tab:Source_properties}}
        \tablehead{& \colhead{Component A} & \colhead{Component B}}
        \startdata
        Right ascension (deg) & $+53.1647632$ & $+53.164735$
        \\
        Declination (deg) & $-27.7746223$ & $-27.774714$
        \\
        NIRCam ID & $130158$ & $130115$
        \\
        NIRSpec ID (1210) & 10014220 & --
        \\
        NIRSpec ID (3215) & 20130158 & --
        \\
        $M_\text{UV} \, (\mathrm{mag})$ & $-19.30 \pm 0.05$ & $-17.43 \pm 0.17$
        \\
        $\beta_\text{UV}$ & $-2.17 \pm 0.03$ & $-2.24 \pm 0.04$
        \\
        $r_e \, (\mathrm{mas})$ & $29 \pm 1$ & $22 \pm 8$
        \\
        $r_e \, (\mathrm{pc})$ & $114 \pm 8$ & $86 \pm 30$
        \\
        $z_\text{spec}$, $z_\text{phot}$ & $11.122_{-0.003}^{+0.005}$ & $10.9 \pm 0.3$
        \enddata
        \tablecomments{
            We distinguish between components A and B (\cref{ssec:Observations:HST_JWST}), for which the coordinates, absolute UV magnitude $M_\text{UV}$, and half-light radii $r_e$ are based on \textsc{forcepho} (B.~D.~Johnson et al. in prep.) fits further detailed in Appendix~\ref{app:Photometric_measurements}. The UV slope $\beta_\text{UV}$ is derived from stellar population synthesis models fit to the photometry (and spectroscopy for component A; \cref{ssec:Discussion:SPS_modelling}). For component A, we list the NIRSpec-derived spectroscopic redshift $z_\text{spec}$ reported by \citet{2024ApJ...976..160H} based on multiple, faint emission lines (combined at $94\%$ confidence; see text for details), whereas for component B the photometric redshift $z_\text{phot}$ derived from NIRCam and MIRI is listed.
        }
    \end{deluxetable}
\endgroup

Motivated by the presence of a nearby faint companion first identified as a potential satellite of \JGSzeleven by \citet{2024ApJ...976..160H}, as shown in \cref{fig:NIRCam_NIRSpec}, our fiducial photometric measurements were obtained with \textsc{forcepho} \citep[B.~D.~Johnson et al. in prep.; see also e.g.][]{2023NatAs...7..611R, 2023ApJ...952...74T, 2025NatAs...9..141B}, which simultaneously constrains the photometry and morphology of both the main galaxy and companion (also respectively labelled components `A' and `B' hereafter). The results are presented in Appendix~\ref{app:Photometric_measurements}, alongside traditional aperture-photometry measurements. Uncertainties on the aperture photometry are measured from a series of random apertures as detailed in \citep{2023ApJS..269...16R}, while \textsc{forcepho} provides a Bayesian forward-modelling approach to estimate uncertainty through standard Markov Chain techniques, which explicitly takes into account the effects of correlated pixel noise, the variation of the NIRCam point spread function (PSF) across photometric bands, neighbouring sources, and uncertainty in the source morphology \citep{2023NatAs...7..611R}. This reveals that the photometry of the companion (B) is statistically fully consistent with $z \approx 11$.

Our spectroscopic data set, meanwhile, consists of a new reduction of the ultra-deep NIRSpec/PRISM spectrum already presented by \citet{2024ApJ...976..160H}. Capturing the main galaxy (A) in the NIRSpec micro-shutter (\cref{fig:NIRCam_NIRSpec}), this consists of $186$ sub-exposures from PIDs 1210 and 3215 with a total exposure time of $75 \, \mathrm{h}$. Here, we exploit the latest reduction pipeline from the NIRSpec GTO Collaboration as described in data release 4 paper II \citep{2025arXiv251001034S}. Largely equivalent to the pipeline behind DR3 \citep{2025ApJS..277....4D}, it adopts more recent NIRSpec calibration files and includes a new correction for intra-shutter offsets. The $186$ sub-exposures are filtered and combined using the prescription described in \citet{2024ApJ...976..160H} and \citet{2025Natur.639..897W}, which is repeated $5000$ times in a bootstrapping procedure to construct the covariance matrix (see Appendix~\ref{app:Covariance_matrix}).

\needspace{2.5cm}
\subsection{ALMA observations}
\label{ssec:Observations:ALMA}

\JGSzeleven was observed in band 6 \citep{2004stt..conf..181E} and 7 \citep{2012ITTST...2...29M} primarily targeting the \OIIIeightyeight line under the A-ranked programme 2023.1.00336.S in Cycle 10 (PIs: Witstok \& Smit). The originally proposed observations employed a spectral tuning with three adjacent spectral windows (SPWs) at the edge of band 6, covering $>99\%$ of the posterior distribution obtained by \citet{2023NatAs...7..622C} on $z_\text{spec, IGM}$ (\cref{sec:Introduction}). Before ALMA observations were taken, however, the refined emission-line redshift reported in \citet{2024ApJ...976..160H} was identified, which would place the FIR \OIII transition with rest-frame wavelength of $88.356394 \, \mathrm{\upmu m}$ at $\nu_\text{obs} = 279.90_{-0.12}^{+0.07} \, \mathrm{GHz}$, narrowly shifting it into band 7 (discussed further in \cref{ssec:Discussion:DLA_absorption}). A request to change the spectral tuning placing four SPWs in band 7 was approved by ALMA (albeit with lowered sensitivity according to the difference in atmospheric transmission), but the band-6 observations accidentally remained in the observing queue and ended up being fully observed.

Observations were carried out between January and November 2024 under variable weather conditions, with precipitable water vapour values of $0.8$-$2.7 \, \mathrm{mm}$ ($0.4$-$1.8 \, \mathrm{mm}$) in band 6 (7). Antennae baselines ranged from $15 \, \mathrm{m}$ up to $314 \, \mathrm{m}$ in band 6, resulting in a reported restoring beam (using Briggs weighting) with a full-width at half maximum (FWHM) of $\ssim 0.8\arcsec \times 0.7\arcsec$ along the major and minor axes, respectively. In band 7, the maximum baseline instead was $1.4 \, \mathrm{km}$, with a beam size of $\ssim 0.4\arcsec \times 0.3\arcsec$. The total on-source time was $19.2 \, \mathrm{h}$, divided across $6 \, \mathrm{h}$ in band 7 and $13.2 \, \mathrm{h}$ in band 6. All data were obtained from the ALMA science archive\footnote{Available at \url{https://almascience.eso.org/asax/} under the project code 2023.1.00336.S.}, calibrated and reduced with the automated pipeline of the Common Astro Software Applications \citep[\textsc{casa};][]{2007ASPC..376..127M} versions 6.5 (band 7) and 6.6 (band 6).
\begin{figure*}
	\centering
	\includegraphics[width=\linewidth]{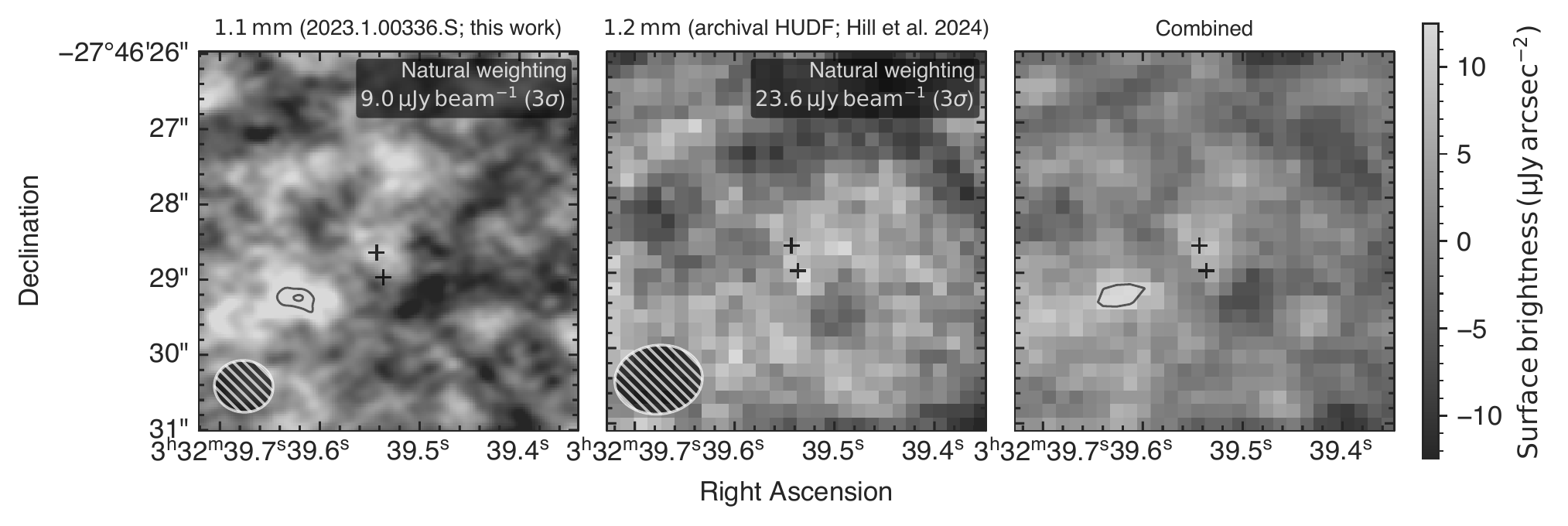}
	\caption{Dust-continuum emission in \JGSzeleven as seen in the observations presented in this work (left), from all archival ALMA imaging in the HUDF \citep[; middle]{2024MNRAS.528.5019H}, and a combination of the two (right; see text for details). Black crosses show the NIRCam-based centroids of the two components (A and B) of \JGSzeleven, a white hatched ellipse shows the ALMA restoring beam. Black contours are drawn from $3\sigma$ with increments of $1\sigma$, not revealing any significant detections at the location of \JGSzeleven.
	}
	\label{fig:Dust-continuum}
\end{figure*}

The mapping from the complex visibility plane onto the image plane was performed with \textsc{casa}'s \textsc{tclean} routine, largely following \citet{2022MNRAS.515.1751W}. We performed the \textsc{uvcontsub} task to subtract the continuum and obtain line-only visibilities, though the underlying dust continuum is not confidently detected (\cref{ssec:Results:Dust_continuum}). We have verified that without the continuum subtraction we obtain indistinguishable results. Data cubes were made with a natural weighting of baselines without applying any tapering to maximise the signal-to-noise ratio (SNR) at the expense of spatial resolution, resulting in a mean beam size of $0.6\arcsec \times 0.4\arcsec$ in band 7. Having explored a range of spectral channel widths, we compromised on rebinning to $\Delta \nu_\text{obs} = 5 \, \mathrm{MHz}$ (corresponding to $\Delta v = 5.4 \, \mathrm{km \, s^{-1}}$ at $279.9 \, \mathrm{GHz}$) to maintain both spectral resolution and SNR.

We created emission-line data cubes only using the band-7 SPWs covering the NIRSpec-based redshift, having verified that there are no other significant, coherent emission-line detections at the spatial location of \JGSzeleven across the entire frequency range probed by the ALMA observations ($269.0$-$292.2 \, \mathrm{GHz}$ across band 6 and 7). Based on initial `dirty' imaging for the first characterisation of the \OIIIeightyeight emission, we made two clean `narrowband' images with frequency widths of $30 \, \mathrm{GHz}$ and $45 \, \mathrm{GHz}$ using the auto-masking continuum cleaning procedure described below. Finally, the data cubes were constructed by cleaning down to a threshold set to two times the root-mean-square (RMS) noise within a custom mask, created from the combination of the two narrowband images.

Continuum images were created by the \textsc{mfs} (multi-frequency synthesis) mode of \textsc{tclean} combining all available observations, stretching across band 6 into band 7 and spanning a total frequency range of $269.0 \, \mathrm{GHz} < \nu_\text{obs} < 292.2 \, \mathrm{GHz}$, except a masked velocity range of $\abs{v} \leq \num{100} \, \mathrm{km/s}$ around the expected location of the \OIIIeightyeight line, given the NIRSpec redshift ($z_\text{spec} = 11.122$). We adopted the mean channel frequency as the representative continuum frequency, $\nu_\text{obs} = 276.7_{-7.7}^{+15.5} \, \mathrm{GHz}$ (errors indicating the full range of the \textsc{mfs} image), which at $z_\text{spec} = 11.122$ translates to a rest-frame wavelength of $\lambda_\text{emit} = 89.4_{-4.8}^{+2.5} \, \mathrm{\upmu m}$. Natural weighting was again used for optimal SNR, which across the entire (band-6 and -7) continuum frequency range resulted in a beam size of $0.9\arcsec \times 0.7\arcsec$. As in \citet{2022MNRAS.515.1751W}, we find the measured RMS noise in naturally weighted images agrees well with the theoretically expected value of the sensitivity of an interferometric image \citep{2017isra.book.....T}. Following \citet{2024MNRAS.535.2068R}, the images were cleaned down to a threshold set to two times the RMS noise, using the auto-masking mode\footnote{See the \textsc{casa} auto-masking guide at \url{https://casaguides.nrao.edu/index.php/Automasking_Guide}.} in \textsc{tclean} (with recommended parameters for long baselines). Flux measurements, both for continuum and line emission, conservatively take into account a $10\%$ systematic flux calibration uncertainty \citep[see Section A.9.2 of the ALMA Proposers' Guide;][]{ALMA_proposers_guide}.

\needspace{2.5cm}
\section{Results}
\label{sec:Results}
\begingroup
    \setlength{\tabcolsep}{6pt} 
    \renewcommand{\arraystretch}{1.25} 
    \begin{deluxetable}{lc}
        \tabletypesize{\footnotesize}
        \tablecaption{FIR properties of \JGSzeleven observed by ALMA\label{tab:FIR_properties}}
        \tablehead{\multicolumn{2}{c}{\OIIIeightyeight line}}
        \startdata
        $t_\text{int} \, (\mathrm{h})$ & $6.0$
        \\
        $\nu_\text{obs, cent.} \, (\mathrm{GHz})$ & $279.901 \pm 0.014$
        \\
        $z_\text{\OIII}$ & $11.1221 \pm 0.0006$
        \\
        $\Delta v \, (\mathrm{km \, s^{-1}})$ & $29 \pm 14$
        \\
        $S_\nu \Delta v \, (\mathrm{mJy \, km \, s^{-1}})$ & $25 \pm 7$
        \\
        $L_\text{\OIIIeightyeight} \, (10^{8} \, \mathrm{L_\odot})$ & $1.1 \pm 0.3$
        \\
        $r_\text{maj} \times r_\text{min} \, (\mathrm{arcsec^2})$ & $0.5_{-0.2}^{+0.2} \times 0.2_{-0.1}^{+0.2}$
        \\
        $r_\text{maj} \times r_\text{min} \, (\mathrm{kpc^2})$ & $1.8_{-0.9}^{+0.8} \times 1.0_{-0.5}^{+0.8}$
        \vspace{0.75ex}
        \\
        \hline\vspace{-2ex}\\
        \multicolumn{2}{c}{Dust continuum}
        \vspace{0.75ex}
        \\
        \hline
        $t_\text{int} \, (\mathrm{h})$ & $19.2$
        \\
        $\lambda_\text{emit} \, (\mathrm{\upmu m})$ & $89.4_{-4.8}^{+2.5}$
        \\
        $S_\nu \, (\mathrm{\upmu Jy})$ & $<9.0$
        \\
        $L_\text{IR} \, (10^{10} \, \mathrm{L_\odot})$ & $\lesssim 3.1$
        \\
        $\text{SFR}_\text{IR} \, (\mathrm{M_\odot \, yr^{-1}})$ & $\lesssim 6$
        \\
        $M_\text{dust} \, (10^{6} \, \mathrm{M_\odot})$ & $\lesssim 1.0$
        \enddata
        \tablecomments{
            Listed properties include the on-source time $t_\text{int}$, observed line frequency $\nu_\text{obs, cent.}$, implied \OIIIeightyeight redshift $z_\text{\OIII}$, line width $\Delta v$, integrated line flux $S_\nu \Delta v$ and line luminosity $L$ (see \cref{ssec:Results:Line_emission} for details), and deconvolved size $r_\text{maj} \times r_\text{min}$, given as the FWHM along major and minor axes (see text for details). Under continuum properties we report rest-frame wavelength $\lambda_\text{emit}$, upper limits (all $3\sigma$) on the flux density $S_\nu$, IR luminosity $L_\text{IR}$, obscured star formation rate $\text{SFR}_\text{IR}$, and dust mass $M_\text{dust}$ (see \cref{ssec:Results:Dust_continuum} for details).
        }
    \end{deluxetable}
\endgroup

\subsection{Dust-continuum emission}
\label{ssec:Results:Dust_continuum}

In \cref{fig:Dust-continuum}, we show the continuum image, which reaches considerable depth combining $19.2 \, \mathrm{h}$ from all available band-6 and -7 observations (\cref{ssec:Observations:ALMA}). It spans a total frequency range of $269.0 \, \mathrm{GHz} < \nu_\text{obs} < 292.2 \, \mathrm{GHz}$ (\cref{ssec:Observations:ALMA}) or rest-frame wavelength of $\lambda_\text{emit} = 89.4_{-4.8}^{+2.5} \, \mathrm{\upmu m}$ at $z = 11.122$. Across $500$ randomly placed, beam-sized apertures we measure a median of $0.2 \, \mathrm{\upmu Jy}$ and standard deviation of $2.8 \, \mathrm{\upmu Jy}$. Still, a beam-sized aperture centred on the NIRCam-based coordinates of \JGSzeleven (i.e. component A) only contains a very faint signal ($1.0\sigma$), from which we conclude we do not robustly detect the continuum emission. This is supported by the slightly shallower archival ALMA imaging in the HUDF \citep{2024MNRAS.528.5019H}, which is also shown in \cref{fig:Dust-continuum} along with an inverse-variance weighted combination of the two maps. Instead, from the data presented in this work we place a $3\sigma$ upper limit of $S_\nu < 9.0 \, \mathrm{\upmu Jy}$ (\cref{tab:FIR_properties}).

Assuming a fiducial dust temperature of $T_\text{dust} = 50 \, \mathrm{K}$ \citep[e.g.][]{2022MNRAS.513.3122S, 2023MNRAS.523.3119W, 2023MNRAS.525.5720J} and a dust-emissivity index of $\beta_\text{IR} = 1.8$ \citep{2023MNRAS.523.3119W}, we use \textsc{mercurius} \citep{2022MNRAS.515.1751W} to infer an infrared (IR) luminosity of $L_\text{IR} \lesssim 3.1 \times 10^{10} \, \mathrm{L_\odot}$ or an obscured star formation rate (SFR) of $\text{SFR}_\text{IR} \lesssim 6 \, \mathrm{M_\odot \, yr^{-1}}$ \citep{2012ARA&A..50..531K}, and a dust mass of $M_\text{dust} \lesssim 1.0 \times 10^{6} \, \mathrm{M_\odot}$. We note that assuming a higher dust temperature would shift the limiting IR luminosity and obscured SFR upwards (less stringent), but lower the upper limit on the dust mass (more stringent). These results will be discussed further in \cref{ssec:Discussion:Dust_content}, and a more in-depth analysis of this non-detection in the context of other high-redshift ALMA continuum observations is presented in T.~J.~L.~C.~Bakx et al. (subm.).

\needspace{2.5cm}
\subsection{Doubly ionised oxygen 88 $\mu m$ fine-structure line}
\label{ssec:Results:Line_emission}

\begin{figure*}
	\centering
	\includegraphics[width=\linewidth]{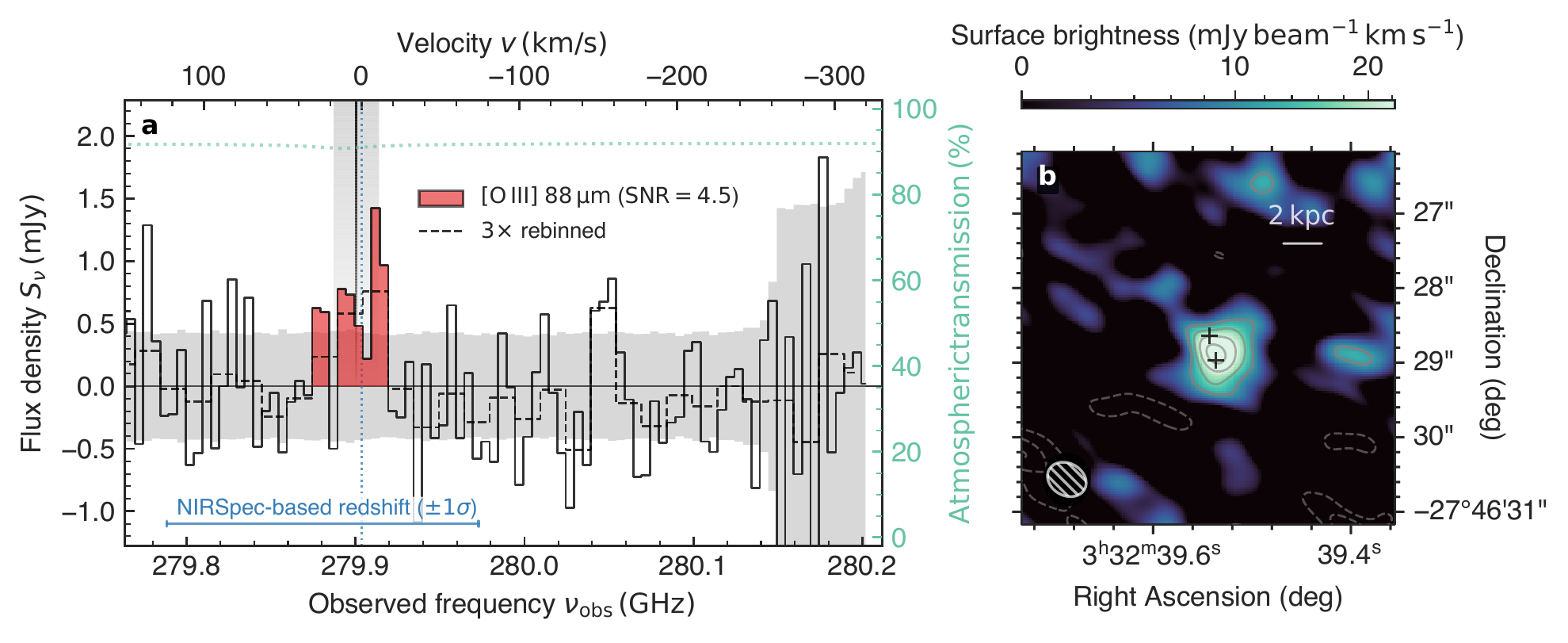}
	\caption{\textbf{a}, Spectrum of \OIIIeightyeight in \JGSzeleven with $\Delta \nu_\text{obs} = 5 \, \mathrm{MHz}$ ($15 \, \mathrm{MHz}$) bins shown by the solid (dashed) black lines (see \cref{ssec:Results:Line_emission} for details). Symmetric grey shading shows $1\sigma$ uncertainty. A vertical dotted blue line shows the NIRSpec-based redshift, with a horizontal range indicating the $\pm 1\sigma$ uncertainty, while the vertical black line and fading grey shading shows the inferred ALMA redshift and corresponding uncertainty. The channel range from which the surface-brightness map is created is coloured in red (\cref{ssec:Results:Line_emission}). The atmospheric transmission is also indicated on the right vertical axis. \textbf{b}, Surface-brightness map of \OIIIeightyeight in \JGSzeleven (\cref{ssec:Results:Line_emission}). Black crosses show the NIRCam-based centroids of the two components (A and B) of \JGSzeleven, a white hatched ellipse shows the ALMA restoring beam. Grey (dashed) contours are drawn at (negative) $2\sigma$, continuing up to $5\sigma$ with increments of $1\sigma$. A horizontal bar indicates a physical scale of $2 \, \mathrm{kpc}$ at $z = 11.122$.
	}
	\label{fig:ALMA_spectrum}
\end{figure*}

Given the spectroscopic redshift based on UV and optical emission lines (\cref{fig:NIRCam_NIRSpec}) first reported in \citet{2024ApJ...976..160H}, \OIIIeightyeight is expected to be located at $\nu_\text{obs} = 279.90_{-0.12}^{+0.07} \, \mathrm{GHz}$. As shown in \cref{fig:ALMA_spectrum}, we indeed find one clear, coherent positive flux excess precisely centred around this frequency. Specifically, we find an excess in channels $32$ through $40$ (marked as the red channels), with velocity offsets of $-14$ to $29 \, \mathrm{km \, s^{-1}}$. We constructed an initial moment-zero map across these channels, from which we extract a spectrum in the central region where $\text{SNR} \geq 2$.\footnote{Uncertainties are empirically measured as the RMS across all two-dimensional pixels (excluding the central $\ssim 2\arcsec \times 2\arcsec$). In the case of the spectrum this is done for each channel separately, and the uncertainty is scaled according to the square root of the number of pixels contained in the aperture.} We repeated the process, now constructing a weighted surface brightness map weighting the contribution of each channel by its spectral flux density, finding convergence is reached after the second iteration. Integrating the resulting spectrum across the full channel range from above (including those with negative flux), we find the integrated signal reaches $4.5\sigma$ significance when applying a weight to the contribution of each spatial pixel in the moment-zero map based on its flux, or $4.1\sigma$ in the unweighted case. The peak SNR in the (un)weighted surface brightness map is $\text{SNR} = 5.5$ (4.2).

To further investigate the integrity of this measurement, we applied the Matched Filtering in 3D (\textsc{mf3d}) algorithm, designed to perform a statistically robust blind search for emission lines in interferometric data \citep{2018ApJ...864...49P}. Briefly, \textsc{mf3d} models the intrinsic emission as a three-dimensional Gaussian distribution in spatial and spectral dimensions, and performs a filtering algorithm in Fourier space. Although the Gaussian approximation may not be a perfect model description in our case (\cref{fig:ALMA_spectrum}), generally speaking the \textsc{mf3d} algorithm is the theoretically optimal detection method \citep[see Appendix~A1 in][]{2018ApJ...864...49P}. First adopting a range of templates with varying spatial and spectral sizes, we find good agreement with the same signal found at a reported $\text{SNR} = 4.14$ with best-matched spatial size of 0 (i.e. a point source) and frequency width of $5$ channels (i.e. $\Delta v = 27 \, \mathrm{km \, s^{-1}}$).

The observed flux excess is decidedly marginal. Nevertheless, both the spatial and spectral locations match the object coordinates and spectroscopic redshift measured a priori by NIRCam and NIRSpec, respectively. Moreover, the significance exceeds the common $3.5\sigma$ threshold for emission-line detections adopted by similar experiments \citep[where spectroscopic redshifts are known in advance; e.g.][]{2020A&A...643A...1L}. Therefore, we consider this a nominal detection of \OIIIeightyeight in \JGSzeleven.

Although the measured flux may marginally overestimated due to \citet{1914smsu.book.....E} bias, at our estimated $\text{SNR} = 4.5$ the magnitude of this effect is likely of the order of $5$-$10\%$ \citep[e.g.][]{2020A&A...643A...2B}, less than or comparable to the systematic uncertainty we include in our flux measurements (\cref{ssec:Observations:ALMA}).

The reputed detection, however, is corroborated by its purity, which we can estimate under the commonly adopted assumption that all negative peaks are purely due to statistical fluctuations \citep{2018ApJ...864...49P, 2024ApJ...964..146F}. Now adopting only templates with intrinsic width of $5$ or $6$ frequency channels to match the putative narrow \OIIIeightyeight line, there are $17$ such narrow negative features found by the \textsc{mf3d} algorithm with $|\text{SNR}| \geq 4.14$ in the full data cube, which spans $19.2\arcsec \times 19.2\arcsec$ on sky across a frequency range $\Delta \nu_\text{obs} = 0.490 \, \mathrm{GHz}$. However, we need to consider that our detection experiment was not a blind search across the entire data cube but was instead carried out over a much smaller effective volume around the NIRSpec-based redshift. Within a search radius of $0.5\arcsec$ over a redshift range of $\Delta z = 0.008$ \citep[i.e. the $\pm 1 \sigma$ uncertainty on the NIRSpec-based redshift;][]{2024ApJ...976..160H} or equivalent frequency range $\Delta \nu_\text{obs} = 1847 \, \mathrm{MHz}$, $>1000\times$ smaller than the full data cube, the expected number of features with $\text{SNR} \geq 4.14$ is then $0.0137$. In other words, given the probability that the observed $\text{SNR} \approx 4.14$ feature is purely due to a statistical fluctuation is approximately $1.4\%$, we estimate a $\ssim 98.6\%$ purity.

We note that due to correlated pixel noise on spatial scales below the beam size \citep{2018ApJ...864...49P}, despite its integrated $\text{SNR} = 4.5$ the purity of this signal is expected to be lower than the nominal $99.9993\%$ probability of a random variable with standard normal distribution to fall between $\pm 4.5\sigma$. Indeed, our purity estimate is in good agreement with the findings of \citet{2020A&A...643A...2B}, who report a $95\%$ purity is reached at a threshold of $\text{SNR} = 3.5$.
\begin{figure*}
	\centering
	\includegraphics[width=\linewidth]{"SEDs_v5.1_extr3_phot_ForcePho_v1.0d.0_bursty-continuity-DM_bpass_v2.3.a+02_CF00_FIR_F_nu"}
	\caption{\textbf{a}, NIRCam false-colour image, where each colour channel was constructed by stacking NIRCam filters as annotated. The placement of the NIRSpec micro-shutters is shown in white (nearly identical across all visits). Dashed coloured circles indicate the $0.3\arcsec$-diameter apertures used to extract the CIRC2 photometry. The \OIIIeightyeight emission and the MIRI/F560W image are overlaid with red and green contours respectively, both drawn at $3\sigma$-$4\sigma$-$5\sigma$. A red (green) hatched ellipse shows the ALMA restoring beam (F560W point spread function). A horizontal bar indicates a physical scale of $1 \, \mathrm{kpc}$ at $z = 11.122$. \textbf{b}, Best-fit model SFHs of the two components, with shading representing $1\sigma$ uncertainty. A solid dark blue line shows the prediction by an `attenuation-free' model \citep[AFM;][]{2024A&A...689A.310F}. The dotted black line and grey shading illustrates the rising SFH prior based on dark matter halo accretion (see text for details), scaled to the stellar mass of component~A. \textbf{c}, SEDs of the main component (A) of \JGSzeleven, as observed by NIRCam, MIRI, and NIRSpec/PRISM (\cref{ssec:Observations:HST_JWST}), as well as the faint neighbouring source (B) whose photometry (filter curves drawn at the bottom) is consistent with having the same redshift. The best-fit \textsc{bagpipes} models are shown as solid (component~A) and dashed (B) coloured lines, with squares indicating the model photometry. \textbf{d}, Magnitude of residuals between observed and modelled flux densities inversely weighted by observational uncertainty, $|\chi|$. \textbf{e}, Agreement between the observed and modelled emission lines and the dust continuum at $88 \, \mathrm{\upmu m}$.
	}
	\label{fig:SEDs}
\end{figure*}

The spectral line shape does not appear to be captured perfectly by a single Gaussian, which could be due to separate components in velocity space. However, due to the limited SNR we are not able at this point to confidently disentangle the spectral peaks (e.g. to link them to distinct spatial components), and therefore conclude these may also be noise artefacts. Instead, we determine a flux-weighted central frequency of $\nu_\text{obs} = 279.901 \pm 0.014 \, \mathrm{GHz}$, which corresponds to a spectroscopic redshift of $z_\text{\OIII} = 11.1221 \pm 0.0006$, in perfect agreement with the redshift measured by NIRSpec from multiple faint emission lines \citep[see \cref{fig:NIRCam_NIRSpec} and][]{2024ApJ...976..160H}. Notably, based on the combined $94\%$ confidence level for the NIRSpec-based redshift solution \citep{2024ApJ...976..160H}, we estimate the joint probability that all observed NIRSpec- and ALMA-observed emission lines coincident at this redshift are purely due to noise is about $(1-0.94) \times 0.0137 = 0.082\%$. We conclude that the redshift solution reported here is robust at a $>99.9\%$ level.

The redshift uncertainty reported here, $\gtrsim \! 5\times$ reduced compared to NIRSpec, is found through a Monte Carlo procedure in which we re-measure these quantities $1000$ times, perturbing each iteration's spectrum with random Gaussian noise scaled according to each spectral bin's uncertainty. Though with considerable uncertainty, we find a line FWHM of $\Delta v = 29 \pm 14 \, \mathrm{km \, s^{-1}}$ from a Gaussian fit to the spectrum, in good agreement with the \textsc{mf3d} estimate. The implications of this relatively narrow line width will be discussed in \cref{ssec:Discussion:SPS_modelling}.

We modelled the moment-zero map as a two-dimensional Gaussian, parametrised by its normalisation, centre, spatial extent along both axes, and position angle (PA), both with and without explicitly performing convolution with the ALMA beam. Sampling the posterior distribution in the convolved case using the \textsc{emcee} package \citep{2013PASP..125..306F}, we find that due to limited SNR, individual parameters are not very well constrained when all freely varied. Bearing in mind this major caveat, there is an indication that the emission is spatially extended: we obtain a lower limit on the deconvolved size along the major axis of $\text{FWHM} > 0.15\arcsec$ or $>0.6 \, \mathrm{kpc}$ (nominally at $95\%$ confidence). Interestingly, the emission furthermore appears to be centred nearer to the companion source: we measure an offset from the main (A) and companion (B) components of $0.3\arcsec$ and $0.1\arcsec$, respectively at $\ssim 2\sigma$ and $\ssim 0.6\sigma$ significance compared to the expected astrometric precision of $0.15\arcsec$ at $\text{SNR} = 4.5$ \citep[cf. Section 10.5.2 of the ALMA Technical Handbook;][]{ALMA_technical_handbook}. This suggests the companion could be at least partly responsible for the observed \OIIIeightyeight emission, which is further supported by the bright, extended MIRI/F560W emission probing $\OIII + \Hbeta$ emission, as will be discussed in \cref{sec:Discussion}.

\needspace{2.5cm}
\section{Discussion}
\label{sec:Discussion}

\subsection{A low-mass, vigorously star-bursting system}
\label{ssec:Discussion:SPS_modelling}

Both components A and B of \JGSzeleven are very compact in the rest-frame UV, yet at the same time the NIRCam imaging indicates they are spatially extended (Appendix~\ref{app:Photometric_measurements}). Since this suggests at most a limited AGN contribution, we inferred physical properties by fitting stellar population synthesis models to the NIRCam measurements of both spatial components using Bayesian Analysis of Galaxies for Physical Inference and Parameter EStimation \citep[\textsc{bagpipes};][]{2018MNRAS.480.4379C}.

Given that the companion photometry is consistent with being at the same redshift as the main galaxy (\cref{ssec:Observations:HST_JWST}), consistent with predictions of frequent merger events at this redshift regime \citep{2025A&A...704A..39K}, in our fiducial fits we let the redshift vary within a $\pm 1\sigma$ confidence interval of the ALMA-based redshift (\cref{ssec:Results:Line_emission}) for both components A and B. In the case of component A, which is captured within the NIRSpec micro-shutter, we simultaneously fit to the ultra-deep NIRSpec/PRISM spectrum (\cref{ssec:Observations:HST_JWST}). In addition, we took into account the (non-detected) $88 \, \mathrm{\upmu m}$ dust-continuum emission (\cref{ssec:Results:Dust_continuum}) into the photometric measurements of component A, which we will show likely contains the bulk of the evolved stellar mass between the two components, and hence is expected to contain the majority of the dust content (if any). Appendix~\ref{app:SPS_modelling} discusses further details of this fitting procedure, which largely follows that of \citet{2025MNRAS.536...27W, 2025Natur.639..897W}.

Based on initial tests, where we found the SFRs of components A and B averaged over the last $10 \, \mathrm{Myr}$ follow a $3:1$ ratio (as further discussed in \cref{ssec:Discussion:Metallicity}), we accordingly distributed the total \OIIIeightyeight luminosity among the main galaxy and companion. These \OIII luminosities, as well as the \OII and \NeIII line fluxes for component A \citep{2024ApJ...976..160H}, were used as an additional direct constraint in our fiducial fitting procedure \citep[see also e.g.][]{2023MNRAS.522.6236T}. The main effect of this is to maintain a level of recent star formation that also better fits the observed flux excess in the MIRI/F560W filter, in which the emission again stretches across both components as shown in \cref{fig:SEDs}a \citep[see also][]{2025A&A...696A..57O}. While this subtle change in the SFH has limited impact on physical parameters such as the stellar mass and SFR averaged over longer timescales ($>10 \, \mathrm{Myr}$) given the modest SNR of these lines, it can make a large difference on photometrically derived redshifts as will be discussed further in \cref{ssec:Discussion:DLA_absorption}. Our best-fit model of component A, which has $16$ parameters and is fit to a total of $18$ photometric bands, $560$ valid wavelength bins, and $3$ emission lines (the \OII doublet being fit as one unresolved line), has a reduced chi-squared value of $\chi_\mathrm{\nu}^2 = 1.62$.
\begingroup
    \setlength{\tabcolsep}{4pt} 
    \renewcommand{\arraystretch}{1.25} 
    \begin{deluxetable}{lcc}
        \tabletypesize{\footnotesize}
        \tablecaption{Physical properties of \JGSzeleven\label{tab:UV_optical_properties}}
        \tablehead{& \colhead{Component A} & \colhead{Component B}}
        \startdata
        \multicolumn{3}{l}{\textit{Inferred physical properties}}
        \\
        $M_* \, (10^{7} \, \mathrm{M_\odot})$ & $27 \pm 4$ & $0.6_{-0.2}^{+0.5}$
        \\
        $\Sigma_* \, (10^2 \, \mathrm{M_\odot \, pc^{-2}})$ & $33_{-4}^{+5}$ & $1.3_{-0.4}^{+1.1}$
        \\
        $Z \, (\mathrm{Z_\odot})$ & $0.37 \pm 0.06$ & $0.09_{-0.03}^{+0.05}$
        \\
        $\text{SFR}_{10} \, (\mathrm{M_\odot \, yr^{-1}})$ & $1.7_{-0.7}^{+1.2}$ & $0.5_{-0.1}^{+0.2}$
        \\
        $\Sigma_\text{SFR, 10} \, (\mathrm{M_\odot \, yr^{-1} \, kpc^{-2}})$ & $21_{-9}^{+14}$ & $12_{-3}^{+5}$
        \\
        $\text{SFR}_{30} \, (\mathrm{M_\odot \, yr^{-1}})$ & $5.1_{-0.7}^{+0.9}$ & $0.20_{-0.06}^{+0.11}$
        \\
        $\Sigma_\text{SFR, 30} \, (\mathrm{M_\odot \, yr^{-1} \, kpc^{-2}})$ & $62_{-9}^{+11}$ & $4.3_{-1.2}^{+2.4}$
        \\
        $t_* \, (\mathrm{Myr})$ & $33_{-4}^{+8}$ & $7_{-4}^{+17}$
        \\
        $A_V \, (\mathrm{mag})$ & $0.11_{-0.05}^{+0.08}$ & $0.16_{-0.06}^{+0.09}$
        \\
        $U_\text{min}$ & $1.5_{-1.1}^{+4.4}$ & --
        \\
        $\gamma$ & $0.2 \pm 0.1$ & --
        \\
        $\log_{10} U$ & $-0.7_{-0.2}^{+0.1}$ & $-1.1_{-0.5}^{+0.4}$
        \\
        $\log_{10} N_\text{\HI} \, (\mathrm{cm^{-2}})$ & $22.4 \pm 0.1$ & $21.7_{-1.5}^{+1.1}$
        \\
        \midrule
        \multicolumn{3}{l}{\textit{Predicted emission-line strengths}$^\dagger$}
        \\
        $F_\text{\Hbeta}$ & $10 \pm 1$ & $8 \pm 1$
        \\
        $F_\mathrm{\OIII \, \lambda \, 5008 \, \Angstrom}$ & $57_{-10}^{+8}$ & $40 \pm 6$
        \\
        $F_\text{\Halpha}$ & $29_{-3}^{+2}$ & $23 \pm 4$
        \\
        $\text{EW}_\mathrm{\text{\OIII} + \Hbeta} \, (\Angstrom)$ & $460_{-85}^{+78}$ & $3170_{-860}^{+670}$
        \\
        $L_\text{\OIIIeightyeight} \, (10^7 \, \mathrm{L_\odot})$ & $8.8_{-0.3}^{+0.3}$ & $2.8_{-0.6}^{+0.7}$
        \\
        $\OIII \, \lambda \, 5008 \, \Angstrom / \text{\OIIIeightyeight}$ & $4.5_{-0.6}^{+0.7}$ & $5.6_{-1.0}^{+0.9}$
        \\
        \midrule
        \multicolumn{3}{l}{\textit{Goodness-of-fit statistics}}
        \\
        $\chi_\text{phot}^2$ ($N_\text{bands}$) & $15.6$ ($18$) & $3.3$ ($17$)
        \\
        $\chi_\text{spec}^2$ ($N_\text{bins}$) & $897$ ($560$) & --
        \\
        $\chi_\text{lines}^2$ ($N_\text{lines}$) & $0.2$ ($3$) & $0.0$ ($1$)
        \\
        $\chi_\nu^2$ ($N_\text{params}$) & $1.62$ $(16)$ & $0.55$ ($12$)
        \enddata
        \tablecomments{
            Listed properties are based on \textsc{bagpipes} fitting to NIRCam, MIRI, NIRSpec$^\ast$, and ALMA observations of components A and B. These include the stellar mass $M_*$ and corresponding surface density $\Sigma_*$, metallicity $Z$, average SFR over the last $10 \, \mathrm{Myr}$ or $30 \, \mathrm{Myr}$ ($\text{SFR}_{10}$ and $\text{SFR}_{30}$) and linked surface densities ($\Sigma_\text{SFR, 10}$ and $\Sigma_\text{SFR, 30}$), mass-weighted age $t_*$, $V$-band attenuation $A_V$, $U_\text{min}$ and $\gamma$ parameters in the \citet{2007ApJ...657..810D} model, ionisation parameter $U$, \HI column density $N_\text{\HI}$, and the predicted strengths (observed flux, rest-frame EWs, and luminosity) of several strong emission lines including \OIIIeightyeight. We report individual $\chi^2$ goodness-of-fit statistics and number of modelled data points for the photometry ($\chi_\text{phot}^2$ and $N_\text{bands}$), spectroscopy ($\chi_\text{spec}^2$ and $N_\text{bins}$), emission lines ($\chi_\text{lines}^2$ and $N_\text{lines}$), and overall reduced $\chi^2$ ($\chi_\text{nu}^2$) normalised by the degrees of freedom (total number of data points minus the number of model parameters, $N_\text{params}$). \\
            $^\ast$ Inferred properties incorporate NIRSpec measurements for component A (component B was outside the micro-shutter). \\
            $^\dagger$ Predicted observed fluxes include the effects of dust obscuration, and are given in units of $10^{-20} \, \mathrm{erg \, s^{-1} \, cm^{-2}}$.
        }
    \end{deluxetable}
\endgroup

Our fiducial fits, which are summarised in \cref{tab:UV_optical_properties} and shown in \cref{fig:SEDs}, employ the latest v2.3 Binary Population and Spectral Synthesis \citep[\textsc{bpass};][]{2022MNRAS.512.5329B} models with an $\alpha$ enhancement of $[ \mathrm{\alpha/Fe} ] = 0.2 \, \mathrm{dex}$. The choice for these $\alpha$-enhanced models is motivated by recent works that have reported observational evidence for the abundances of $\alpha$ elements (O, Ne, Mg, etc.) to become enhanced relative to iron at higher redshift \citep[$3\times$ solar at $z \gtrsim 3$;][]{2021MNRAS.505..903C, 2024ApJ...966..234B, 2024MNRAS.532.3102S, 2025ApJ...994..165P, 2025ApJ...983L..30B}. This is in line with predictions from chemical evolution models, in which type-II supernovae (SNe) begin releasing large amounts of oxygen into the ISM within as little as $3 \, \mathrm{Myr}$ after a burst of star formation, whereas the majority of iron is produced by type-Ia SNe on timescales longer than $100 \, \mathrm{Myr}$ \citep{2019A&ARv..27....3M, 2020ApJ...900..179K}.

By contrast, \citet{2025ApJ...994...65N} recently argued that \JGSzeleven may be \emph{enhanced} in iron, prompting us to also explore models with a decreased $\alpha$-to-iron ratio of $[ \mathrm{\alpha/Fe} ] = -0.2 \, \mathrm{dex}$ (the lowest available value in \textsc{bpass} v2.3). However, we find this does not noticeably improve the overall fit compared to the fiducial model, neither in photometry or spectroscopy with respective goodness-of-fit statistics of $\chi_\text{phot}^2 = 16.0$ versus $\chi_\text{phot}^2 = 15.6$, and $\chi_\text{spec}^2 = 902$ versus $\chi_\text{spec}^2 = 897$. Though in apparent disagreement with \citet{2025ApJ...994...65N}, it should be noted that their measured $\alpha$-to-iron ratio of $[ \mathrm{\alpha/Fe} ] = -0.91 \, \mathrm{dex}$ is still $5\times$ lower than the value considered here.
\begin{figure}
	\centering
	\includegraphics[width=\linewidth]{"Velocity_dispersion"}
	\caption{Stellar velocity dispersion ($\sigma_*$) as function of stellar mass ($M_*$). Literature measurements of the $z > 11$ galaxies \JGSzeleven (this work), GHZ2 \citep{2024ApJ...977L...9Z}, and JADES-GS-z14-0 \citep{2025ApJ...988...19S, 2025A&A...696A..87C} are indicated by black points. We estimated their stellar velocity dispersion by converting the measured \OIIIeightyeight line width as in \citet[; see \cref{ssec:Discussion:SPS_modelling} for details]{2023A&A...677A.145U}. The stellar mass of \JGSzeleven shown here stretches across the best-fit value presented in this work (\cref{tab:Source_properties}) and the value presented in \citet{2024ApJ...976..160H} to illustrate the expected degree of systematic uncertainty. Nearby systems from the compilation in \citet{2014MNRAS.443.1151N} are shown by coloured points. This sample includes elliptical and S0 galaxies (Es/S0s), compact ellipticals (cEs), dwarf ellipticals and S0 galaxies (dEs/dS0s), ultra-compact dwarfs (UCDs), globular clusters (GCs), and dwarf spheroidals (dSphs).
	}
	\label{fig:Velocity_dispersion}
\end{figure}

As for the star formation history (SFH), we adopted the physically motivated rising SFH prior proposed by \citet{2025MNRAS.537.1826T}. As in \citet{2025ApJ...992..212W}, however, we adopt a $\text{SFR}(z) \propto (1+z)^{-4.5}$ scaling based on halo mass abundance matching in the \textsc{abacus} $N$-body simulations \citep{2021MNRAS.508.4017M, 2024Natur.633..318C} instead of the \citet{2013MNRAS.435..999D} analytic approximation of dark matter halo accretion rates, which focussed on a lower-redshift regime than considered here. For the main galaxy (A) we find a stellar mass of $M_* = (2.7 \pm 0.4) \times 10^{8} \, \mathrm{M_\odot}$, somewhat lower than initial NIRCam- \citep{2023NatAs...7..611R} and NIRSpec-based \citep{2023NatAs...7..622C} analyses, but in good agreement with the more recent findings of \citet{2024ApJ...976..160H}. We do find a higher SFR and younger age compared to all these works, also without directly fitting to emission-line strengths.

Interestingly, the best-fit model for component A rejects the rising-SFH prior in preferring a slight downturn in SFR in the last $\ssim 40 \, \mathrm{Myr}$, which results in a mild Balmer break favoured by the redder NIRCam bands, the red end of the NIRSpec/PRISM coverage, and the MIRI/F770W filter that is mainly sensitive to the continuum. We note that this agrees with predictions from the `attenuation-free' model \citep{2024A&A...689A.310F}, in which the existence of blue monsters is explained by powerful radiation-driven outflows evacuating most of the dust content, followed by a post-starburst phase. Nevertheless, the break is not as pronounced as in `mini-quenched' galaxies \citep{2023ApJ...949L..23S, 2024Natur.629...53L, 2025A&A...697A..88L, 2025MNRAS.537..112W, 2025MNRAS.537.3662T, 2025A&A...697A..90B} and our models indicate the observed $\ssim 0.4 \, \mathrm{mag}$ excess in the MIRI/F560W filter---which is contaminated by the $\OIII + \Hbeta$ complex---is likely due not only to the Balmer break but also to moderately strong line emission from ongoing star formation.

As indeed expected in a merger-induced starburst scenario \citep{2025A&A...704A..39K}, the photometry of the fainter companion similarly indicates a rapidly rising SFH with recent ($10 \, \mathrm{Myr}$) SFR nearly a third of that of the main galaxy even though it is subdominant in stellar mass by $\ssim 40\times$, which again leads to a MIRI/F560W excess from extremely strong line emission, $\text{EW}_\mathrm{\text{\OIII} + \Hbeta} > 3000 \, \Angstrom$ \citep[cf.][]{2024MNRAS.533.1111E, 2024MNRAS.535.1796B}. As such, the MIRI coverage of the rest-frame optical redshifted beyond $5 \, \mathrm{\upmu m}$ proves a powerful diagnostic of the stellar populations and ISM conditions even at $z > 10$ \citep[see also][]{2023ApJ...952..143R, 2024ApJ...969...12R, 2025NatAs...9..155Z, 2025NatAs...9..729H, 2025A&A...695A.250A}.

In the following, we will place the relatively small inferred stellar masses in the context of the apparently narrow \OIIIeightyeight line. In \cref{fig:Velocity_dispersion}, the measured line width of \JGSzeleven is compared against the compilation of measurements in \citet{2014MNRAS.443.1151N} for local galaxies and globular clusters (GCs) as a function of their stellar mass. This local-galaxy sequence stretches across more than seven orders of magnitude in stellar mass and includes elliptical and S0 galaxies, compact ellipticals, dwarf ellipticals and S0 galaxies, ultra-compact dwarfs, GCs, and dwarf spheroidals. For the three $z > 11$ galaxies with ALMA-detected \OIIIeightyeight lines---GHZ2 at $z = 12.33$ \citep{2024ApJ...977L...9Z}, JADES-GS-z14-0 at $z = 14.18$ \citep{2025ApJ...988...19S, 2025A&A...696A..87C}, and \JGSzeleven---we estimate the integrated stellar velocity dispersion $\sigma_*$ by applying small ($<0.2 \, \mathrm{dex}$) corrections to the measured ionised gas velocity dispersion, based on observed trends between stellar and ionised gas kinematics in $z \sim 1$ galaxies \citep{2018ApJ...868L..36B}.

All three $z > 11$ systems fall into the stellar-mass regime populated by dwarf ellipticals, dwarf S0 galaxies, and ultra-compact dwarfs. Whereas GHZ2 and JADES-GS-z14-0 land towards the upper end of the envelope (higher $\sigma_*$), \JGSzeleven is located at the lower end of (but consistent with) the local sequence, though we again emphasise the considerable uncertainty as discussed in \cref{ssec:Results:Line_emission}.

Following \citet{2025A&A...696A..87C}, we tentatively estimate the dynamical mass through the approach of \citet{2023A&A...677A.145U},
\begin{equation}
    \label{eq:Dynamical_mass}
    M_\text{dyn} = K(q) K(n) \, \frac{\sigma_*^2 R_e}{G} \, ,
\end{equation}
where $G$ is the gravitational constant, $\sigma_*$ the integrated stellar velocity dispersion, $R_e$ the effective radius, $K(n) = 8.87 - 0.831n + 0.0241n^2$ is a correction based on the \citet{1963BAAA....6...41S} index \citep{2006MNRAS.366.1126C} and $K(q) = ( 0.87 + 0.38e^{-3.71(1-q)} )^2$ similarly takes into account for the axis ratio $q$ \citep{2022ApJ...936....9V}. The stellar velocity dispersion $\sigma_*$ is estimated based on the ionised gas velocity dispersion as described above.

This yields $M_\text{dyn} \approx (9.3 \pm 8.3 \, (\mathrm{stat.}) \pm 4.8 \, (\mathrm{syst.})) \times 10^{8} \, \mathrm{M_\odot}$ based on the estimated circularised deconvolved radius of the \OIII emission (\cref{ssec:Results:Line_emission}), with systematic uncertainty estimated from a wide range of possible \citeauthor{1963BAAA....6...41S} indices ($0.5 < n < 4$) and axis ratios ($0.3 < q < 1$). A mild tension with previous stellar mass estimates approaching $M_* \approx 10^9 \, \mathrm{M_\odot}$ \citep{2023NatAs...7..611R} is relaxed by our reduced estimate of the combined stellar mass of the system, $M_* \approx 3 \times 10^8 \, \mathrm{M_\odot}$, perhaps even hinting at a dark-matter dominated system as seen and predicted in low-mass galaxies at an early evolutionary stage \citep{2021A&A...653A..20S, 2024ApJ...967L..40D, 2024A&A...684A..87D, 2026MNRAS.545f2092M}.

However, as also reflected by the large statistical uncertainties, we should caveat that the low SNR of our line detection almost certainly hinders an accurate measurement of its full spatial extent and spectral width, which will require deeper observations to be conclusively established. Moreover, the \OIII emission originates in highly ionised gas likely tracing central star-forming regions, such that it is unclear whether its spectral width can be used as a reliable estimator of the total dynamical mass \citep[e.g.][]{2025ApJ...987L...2H}.

\needspace{2.5cm}
\subsection{Neutral atomic gas puts a damper on distant Lyman-break galaxies}
\label{ssec:Discussion:DLA_absorption}

In the SED modelling of the main component where we simultaneously fit to the deep NIRSpec/PRISM spectrum and photometry, the inclusion of a proximate DLA absorber (Appendix~\ref{app:SPS_modelling}) independently recovers the \HI column density inferred by \citet{2024ApJ...976..160H} within $1\sigma$ uncertainties. At $\log_{10} N_\text{\HI} \, (\mathrm{cm^{-2}}) = 22.4 \pm 0.1$, \JGSzeleven is among a growing number of $z \gtrsim 9$ sources \citep[e.g.][]{2024Sci...384..890H, 2025A&A...696A..87C} where excessive DLA absorption---with \HI column densities over two orders of magnitude larger than the classical DLA threshold \citep[i.e. $N_\text{\HI} > 2 \times 10^{20} \, \mathrm{cm^{-2}}$;][]{2005ARA&A..43..861W}---is required to reconcile the \Lya break with precise emission-line redshifts.

\begin{figure}
	\centering
	\includegraphics[width=\linewidth]{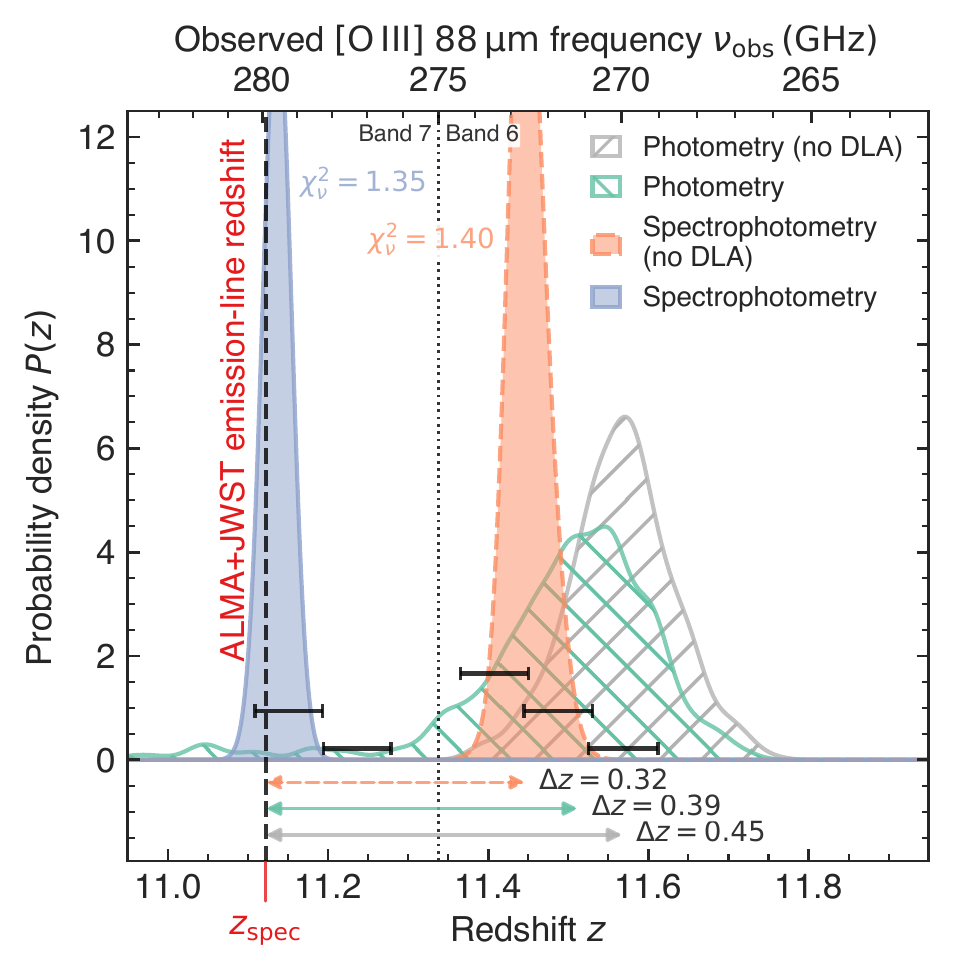}
	\caption{Redshift distributions of the main component (A) under different SED modelling assumptions, both with (`spectrophotometry'; coloured shading) and without (`photometry'; hashed) simultaneously fitting to the NIRSpec/PRISM spectrum, as well as with and without including DLA absorption (different shading and line styles according to the legend). The band-6 and -7 frequency coverage of ALMA is shown by black horizontal lines representing individual SPWs (\cref{ssec:Observations:ALMA}). The true spectroscopic redshift based on the combination of ALMA- and JWST-detected emission lines, shown by the vertical black dashed line, is only recovered by models allowing for DLA absorption. Without considering DLA absorption, all cases (significantly) overestimate the true redshift even when fitting to spectroscopic data.
	}
	\label{fig:Redshift_distributions}
\end{figure}

This is illustrated in \cref{fig:Redshift_distributions}, where we show the posterior distributions of the redshift of the main galaxy (A) obtained with four variants of our SED modelling approach: both with and without simultaneously fitting to the PRISM spectrum, and with and without taking into account DLA absorption. The posteriors are compared to the joint ALMA and JWST spectroscopic redshift that is virtually free of systematic uncertainty mainly thanks to ALMA, which for $5 \, \mathrm{MHz}$ channels at $\nu_\text{obs} = 280 \, \mathrm{GHz}$ has a spectral resolution of $R \approx \num{50000}$ \citep[cf. Section 5.5.2 of the ALMA Technical Handbook;][]{ALMA_technical_handbook}.

Without DLA absorption, the redshift is significantly overestimated when fitting to photometric NIRCam and MIRI data ($\Delta z = 0.44$), as first became clear from the analysis by \citet{2024ApJ...976..160H}. While including DLA absorption in the photometric redshift estimate does allow for the possibility of lower-redshift solutions, there is still a strong bias compared to the true spectroscopic redshift ($\Delta z = 0.40$). Interestingly, we find that our a-posteriori knowledge of the SFH and metallicity as encoded by the presence of a bright \OIIIeightyeight line is able to largely resolve this bias. When including its strength as an additional constraint\footnote{Similar to our fiducial spectrophotometric fits described in \cref{ssec:Discussion:SPS_modelling}, though we do not include \OII or \NeIII here.}, the redshift estimate as well as the fit itself improves from $z_\text{phot} = 11.5 \pm 0.1$ ($\chi_\nu^2 = 1.46$) to $10.8_{-0.3}^{+0.4}$ ($\chi_\nu^2 = 1.18$; distribution not shown here). This demonstrates the importance of the priors as well as the information contained purely in the \OIIIeightyeight line strength.

A similar overestimation occurs even when fitting to the spectroscopic data ($\Delta z = 0.29$). In this approach, the faint \CIV, \OII, and \NeIII emission lines (shown in \cref{fig:NIRCam_NIRSpec}) are not distinctive enough to pinpoint the redshift. Despite the $75 \, \mathrm{h}$ depth of the PRISM spectrum (\cref{ssec:Observations:HST_JWST}) these lines are all detected at $\lesssim 3\sigma$ individually, and only reveal a single, statistically robust redshift solution at $z_\text{spec} = 11.122_{-0.003}^{+0.005}$ through a careful `redshift sweep' procedure (independent of the \Lya break) as detailed in \citet{2024ApJ...976..160H}.

When fitting to the entire spectrum as done here, we instead find the constraints on the redshift are principally driven by the \Lya break. In the spectroscopic case, we do find the inclusion of a DLA component does marginally improve the goodness of fit, which suggests that at sufficiently high SNR, the combination of faint emission lines and the shape of the \Lya break is able to inform us on the presence of DLA absorption. Even in the best-case scenario, fitting the spectroscopic data with potential DLA absorption taken into account, we find an offset, although very minor \citep[smaller than a single PRISM wavelength bin, $\Delta \lambda_\text{obs} \approx 0.012 \, \mathrm{\upmu m}$ or $\Delta z \approx 0.1$ at $\lambda_\text{obs} = 1.5 \, \mathrm{\upmu m}$;][]{2022A&A...661A..80J} that is likely due to residual wavelength calibration uncertainties \citep{2024A&A...690A.288B, 2025ApJS..277....4D, 2025arXiv251001034S}. IGM damping-wing absorption, which is not explicitly accounted for here since we know it to be overshadowed by the remarkably strong DLA absorption, could in general further reduce the redshift bias, though one would need to marginalise over the many uncertainties in ionised bubble size and mean IGM neutral fraction \citep[e.g.][]{2024MNRAS.531L..34K, 2025A&A...697A..89C, 2025arXiv250111702M, 2025ApJ...987...82H}.

The seemingly extreme DLA absorption, however, may not be entirely unexpected given the compactness of \JGSzeleven, a common feature of $z \gtrsim 9$ galaxies \citep{2023NatAs...7..611R, 2023ApJ...949L..34H, 2023ApJ...952...74T, 2024Natur.633..318C, 2025ApJ...980..138H, 2025Natur.639..897W, 2025arXiv250511263N}. For a given neutral gas mass $M_\text{\HI}$, the column density $N_\text{HI}$ increases quadratically with the inverse of the effective radius, $N_\text{\HI} \propto r^{-2}$. Following \citet{2025A&A...693A..60H}, under the assumption that the absorption occurs predominantly within an extended neutral gas reservoir of the main galaxy (A) we estimate the atomic gas mass to be $M_\text{\HI} = 1.5 \pm 0.3 \times 10^{8} \, \mathrm{M_\odot}$. This assumes the neutral gas disc extends out to three times the UV size (with half-light radius of $r_e \approx 100 \, \mathrm{pc}$; \cref{tab:Source_properties}), motivated by ALMA observations of $z \sim 6$ galaxies \citep[e.g.][]{2020ApJ...900....1F, 2021A&A...649A..31H, 2022ApJ...934..144F}. Assuming that cold, \HI gas dominates the gas budget, this implies a modest gas fraction $f_\text{gas} = M_\text{gas}/(M_* + M_\text{gas}) = 35 \pm 6\%$, which is even on the lower end of expectations for metal-poor galaxies \citep[e.g.][]{2019A&A...623A...5D}.

Our estimated SFR surface density of $\Sigma_\text{SFR, 30} = 62_{-9}^{+11} \, \mathrm{M_\odot \, yr^{-1} \, kpc^{-2}}$ (\cref{tab:UV_optical_properties}) and implied (atomic) gas mass surface density of $\Sigma_\text{gas} \approx 200 \, \mathrm{M_\odot \, pc^{-2}}$ would then place \JGSzeleven approximately two orders of magnitude above the Kennicutt-Schmidt (KS) relation, in a highly starburst regime that may be commonplace in the early Universe \citep{2024MNRAS.527...10V}. In the local Universe, this regime is also occupied by metal-poor starburst galaxies \citep{2021ApJ...908...61K}, although there are indications that dwarf galaxies possess large reservoirs of extended \HI and CO-dark molecular gas that could bring these systems closer to the KS relation \citep{2024ARA&A..62..113H}. Similarly, the total gas mass (and hence the gas fraction) of \JGSzeleven may well be larger, since a substantial fraction of the cold gas should be in the molecular phase at such high density \citep[as also pointed out by][]{2024A&A...689A.152D}. On the other hand, we note that neutral, atomic gas outside of the ISM of component A---potentially associated with the companion, if it is situated in front of the main galaxy---could be (partly) responsible for the DLA absorption, which would instead bring down the gas mass surface density.

\needspace{2.5cm}
\subsection{Do very early galaxies already contain dust?}
\label{ssec:Discussion:Dust_content}

First of all, we compare the stellar mass in the context of our limits on the dust mass ($M_\text{dust} \lesssim 1.0 \times 10^{6} \, \mathrm{M_\odot}$; \cref{ssec:Results:Dust_continuum}), which nominally yields a mass ratio $M_\text{dust} / M_* \lesssim 0.4\%$. This ratio is consistent with dust production by SNe regardless of whether destruction takes place in the reverse shock \citep{2018ApJ...868...62G, 2024A&ARv..32....2S}. Still, it implies that grain growth, which is thought to be required for the rapid observed build-up of dust masses at $z \lesssim 8$ \citep{2023MNRAS.519.4632D, 2025ApJ...982....7N}, is not yet very effective \citep{2025arXiv250119384M}. We caution, however, that many caveats should be taken into account considering the estimation of both masses, on the one hand (regarding stellar mass) including variations in the SFH and IMF \citep[e.g.][]{2022ApJ...927..170T, 2023MNRAS.519.5859W, 2024A&A...686A.138C}, and on the other hand the dust temperature, emissivity index $\beta_\text{IR}$, and absorption cross section \citep[e.g.][]{2022MNRAS.515.1751W, 2023MNRAS.523.3119W}.

Another way to explore the potential nature of the absorbing gas is to test whether the observed UV properties can be reconciled with the estimated dust content. First of all, the limiting IR luminosity implies an IR excess (IRX) of $L_\text{IR}/L_\text{UV} \lesssim 2.6$, which given the UV slope of \JGSzeleven, $\beta_\text{UV} \approx -2.2$ \citep[\cref{tab:Source_properties}; see also][]{2024ApJ...976..160H}, is fully consistent with the IRX-$\beta_\text{UV}$ relation observed at $z \sim 7$ \citep{2024MNRAS.527.5808B}. The dust-to-gas ratio of $\lesssim 1\%$ would similarly be in agreement with observations at sub-solar metallicity \citep{2019A&A...623A...5D}. Despite our deep continuum measurements, however, the inferred attenuation in the rest-frame UV ($A_V \approx 0.1 \, \mathrm{mag}$; \cref{tab:UV_optical_properties}) in this case is still a more constraining dust probe than the non-detected FIR dust emission. Following \citet{2023MNRAS.520.2445Z} and \citet{2024A&A...689A.310F}, we estimate that such a seemingly low attenuation would only require a dust mass of $M_\text{dust} \approx 10^{3} \, \mathrm{M_\odot}$ if it were compressed within the measured (UV) size of \JGSzeleven, $r_e \approx 100 \, \mathrm{pc}$, three orders of magnitude below our upper limit on the dust mass (\cref{ssec:Results:Dust_continuum}).

Therefore, if a considerable amount of dust grains has already been produced by SNe, one or several of the following effects should occur: (i) the spatial distribution of the dust is patchy such that we are seeing largely unobscured regions \citep[as has indeed been observed directly at $z > 6$; e.g.][]{2017A&A...605A..42C, 2018MNRAS.481.1631B, 2022MNRAS.515.3126I, 2023ApJ...952....9T}, (ii) the dust has been expelled \citep[out to kiloparsec scales;][]{2025A&A...694A.215F}, or (iii) a combination of a different grain composition, size distribution, and/or dust-star geometry gives rise to a grey attenuation law that does not noticeably redden the UV \citep[as may be expected for pure SN dust;][]{2025ApJ...985L..21M}.

We now turn to the consistency between the dust attenuation and \HI column density by exploiting well-established scaling relations between the gas, metal, and dust content of the diffuse ISM. The diffuse ISM in the Milky Way, where most gas is in the form of \HI, follows a conversion between gas column density and V-band attenuation of $N_\text{H}/A_V = (2.2_{-0.4}^{+0.3}) \times 10^{21} \, \mathrm{cm^{-2} \, mag^{-1}}$ \citep{2011A&A...533A..16W}. Therefore, our inferred column density of $N_\text{HI} = 2.6_{-0.4}^{+0.5} \times 10^{22} \, \mathrm{cm^{-2}}$ would imply $A_V \approx 10 \, \mathrm{mag}$ at solar metallicity. At fixed dust-to-metal ratio (DTM), this in turn means that the absorbing gas should have a metallicity of approximately $1\%$ solar to be reconciled with the observed $A_V \approx 0.1 \, \mathrm{mag}$. However, observations show a distinct decrease in the DTM towards lower metallicities \citep{2024A&A...681A..64K}, so the metallicity of the foreground HI is likely even lower. Because this is significantly lower than both the SED-derived value and the expectation based on the \OIIIeightyeight luminosity (\cref{ssec:Discussion:Metallicity}), it seems plausible that the absorbing gas is in fact a separate, more chemically pristine (or at least dust-poor) medium, perhaps representing an extended neutral gas disc surrounding the UV-bright star-forming regions in \JGSzeleven \citep[as also suggested for JADES-GS-z14-0;][]{2025ApJ...987L...2H}.

\needspace{2.5cm}
\subsection{Drivers of strong oxygen emission}
\label{ssec:Discussion:Metallicity}
\begin{figure}
	\centering
	\includegraphics[width=\linewidth]{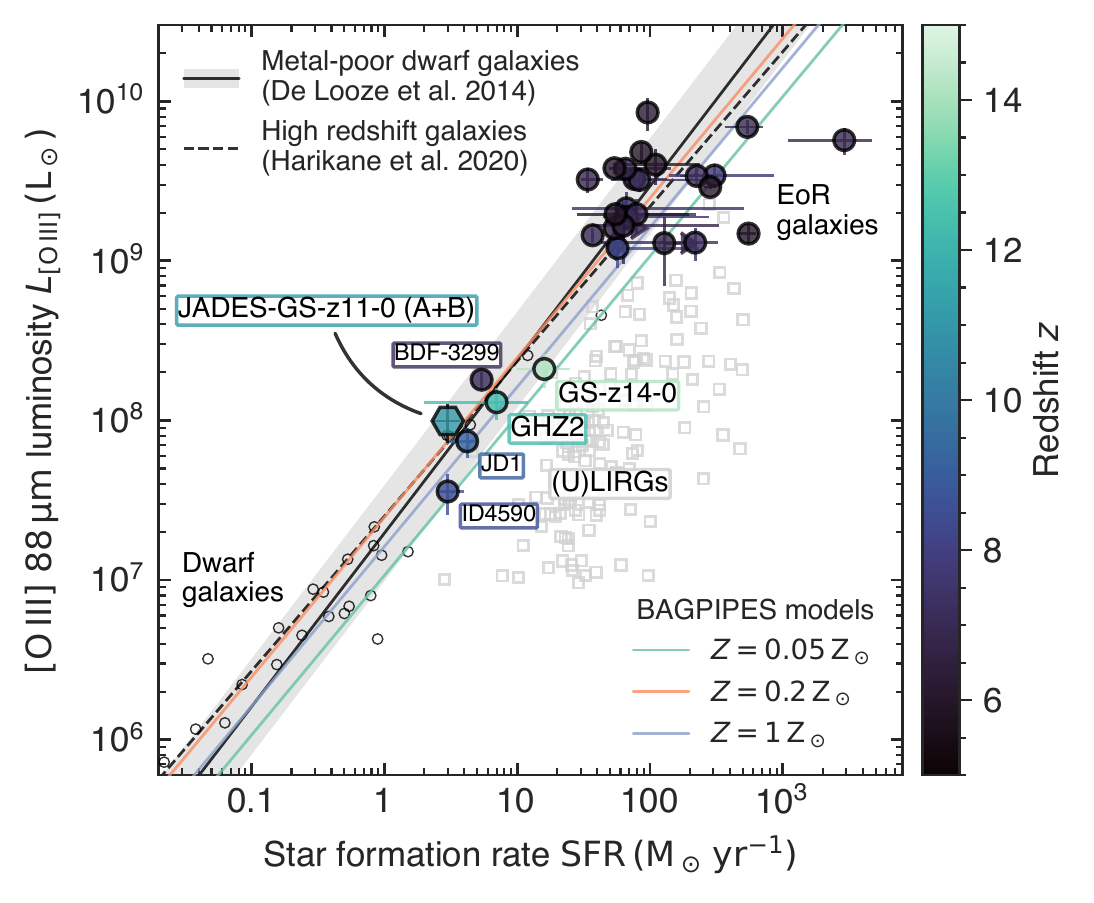}
	\caption{\OIIIeightyeight luminosity versus SFR. We show several possibilities for \JGSzeleven: a hexagon shows the combined SFR of components A and B averaged over $10 \, \mathrm{Myr}$. EoR galaxies with \OIIIeightyeight detections, compiled by \citet{2024MNRAS.532.2270B} and \citet[, in prep.]{2024MNRAS.527.6867A}, are shown as circles colour-coded by redshift. BDF-3299 \citep{2017A&A...605A..42C}, ID4590 \citep{2024ApJ...964..146F}, MACS1149-JD1 \citep{2018Natur.557..392H}, GHZ2 \citep{2024ApJ...977L...9Z}, and JADES-GS-z14-0 \citep{2025ApJ...988...19S, 2025A&A...696A..87C} are labelled individually (by shorthand). Samples of local galaxies are shown as open symbols, with (U)LIRGs \citep{2017ApJ...846...32D} shown as grey squares and local metal-poor dwarf galaxies \citep{2015A&A...578A..53C, 2019A&A...626A..23C} as small black circles. The black solid line shows the \citet{2014A&A...568A..62D} fit to this population, while the dashed black line is the fit to high-redshift galaxies by \citet{2020ApJ...896...93H}. Coloured lines show standard \textsc{Cloudy} models incorporated in \textsc{bagpipes}, for a burst of star formation aged $1 \, \mathrm{Myr}$ (dashed lines) or $10 \, \mathrm{Myr}$ (solid lines) and metallicities of $5\%$, $20\%$ and $50\%$ solar (coloured according to the legend).
	}
	\label{fig:L_OIII_SFR}
\end{figure}

Recent ALMA observations of two of the brightest $z > 12$ galaxies have confirmed that the build-up of elements heavier than helium happens both very early and quickly. Less than $350 \, \mathrm{Myr}$ after the Big Bang, GHZ2 \citep{2024ApJ...977L...9Z} and JADES-GS-z14-0 \citep{2025ApJ...988...19S, 2025A&A...696A..87C} are already found to have oxygen levels of the order of $10\%$ of the solar abundance. Remarkably, they land perfectly within the scatter of the \citet{2014A&A...568A..62D} relation between \OIIIeightyeight luminosity and SFR calibrated on local, metal-poor dwarf galaxies. This relation is shown in \cref{fig:L_OIII_SFR}, highlighting that these $z \gtrsim 11$ galaxies occupy a locus of intermediate SFR in between local metal-poor dwarf galaxies \citep{2015A&A...578A..53C, 2019A&A...626A..23C} and bright, \OIIIeightyeight-detected reionisation-era galaxies as compiled by \citet{2024MNRAS.532.2270B} and \citet[, in prep.]{2024MNRAS.527.6867A}. All of these exhibit elevated \OIIIeightyeight luminosity at fixed SFR compared to (ultra-)luminous infrared galaxies or (U)LIRGs \citep{2017ApJ...846...32D}.

If we assume the \citet{2014A&A...568A..62D} relation to hold in the very early Universe, based on the combined SFR between components A and B averaged over the last $10 \, \mathrm{Myr}$, $\text{SFR}_{10} = 2.3 \pm 0.8 \, \mathrm{M_\odot \, yr^{-1}}$, we would expect \JGSzeleven to have a total \OIIIeightyeight luminosity of $L_\text{\OIII} = (0.5 \pm 0.2) \times 10^8 \, \mathrm{L_\odot}$. Taking into account measurement uncertainties and the systematic $0.3 \, \mathrm{dex}$ scatter in this relation \citep{2014A&A...568A..62D}, this estimate is in agreement with our measured value of $L_\text{\OIII} = (1.1 \pm 0.3) \times 10^8 \, \mathrm{L_\odot}$. While admittedly there is a degree of circularity in this argument, since the \OIII luminosity was used as input to the SED fitting (\cref{ssec:Discussion:SPS_modelling}), we note that this is only one in a total of $18+560+3=581$ data points that were fit to (which was furthermore not perfectly reproduced), and we obtained similar constraints on the SFR when leaving out the \OIII luminosity.

Altogether, this suggests the ISM of \JGSzeleven is enriched to a similar level as GHZ2 and JADES-GS-z14-0 even if it is fainter in the UV by factors of respectively $3$ and $4$. Adopting the \OIIIeightyeight metallicity calibration from \citet{2020ApJ...903..150J}, we obtain an oxygen abundance\footnote{Or more specifically, the O$^{2+}$ abundance. This therefore strictly represents a lower limit on the total oxygen abundance, although $\text{O}^{2+}$ typically dominates the oxygen budget at low metallicity \citep{2017MNRAS.465.1384C}.} of $12 + \log(\mathrm{O/H}) = 8.07 \pm 0.13$. This implies a gas-phase metallicity of $Z = 0.24 \pm 0.07 \, \mathrm{Z_\odot}$ assuming a solar abundance of $12 + \log(\mathrm{O/H})_\odot = 8.69$ \citep{2009ARA&A..47..481A}, in agreement with our SED-based metallicity estimate (\cref{ssec:Discussion:SPS_modelling}).

Interestingly, even if the companion (B) is several times fainter than the main galaxy (A) in the rest-frame UV, its steeply rising SFH suggested by the excess seen in the MIRI/F560W filter causes its $\text{SFR}_{10}$ to be roughly a third of the main galaxy (\cref{tab:UV_optical_properties}). In this case, it would provide a similarly significant contribution to the \OIIIeightyeight emission, with a predicted $L_\text{\OIIIeightyeight} = 2.8_{-0.6}^{+0.7} \times 10^7 \, \mathrm{L_\odot}$. Combined with the apparently spatially extended MIRI/F560W and \OIIIeightyeight emission centred in between the two components, this supports the scenario where the \OIIIeightyeight emission originates in both components of \JGSzeleven.

For further context, we compare our results with the predictions from simple \textsc{Cloudy} models \citep{2017RMxAA..53..385F, 2023RMxAA..59..327C}. Here, we consider the \textsc{bagpipes} implementation of \textsc{Cloudy} using the v2.3 \textsc{bpass} models, again adopting a fiducial $\alpha$ enhancement\footnote{We note that at fixed metallicity, the $\alpha$-enhanced v2.3 \textsc{bpass} models do not result in significantly different ionising outputs \citep{2022MNRAS.512.5329B} and therefore share largely similar predicted nebular emission spectra.} of $[ \mathrm{\alpha/Fe} ] = 0.2 \, \mathrm{dex}$ (cf. Appendix~\ref{app:SPS_modelling}). This follows the \citet{2017ApJ...840...44B} prescription for modelling nebular continuum and line emission using an ionisation-bounded, spherical \HII region of constant density ($n_\text{H} = 100 \, \mathrm{cm^{-3}}$) illuminated by a single stellar population (SSP) of given age and metallicity for a specified ionisation parameter $U$ \citep[see also][]{2018MNRAS.480.4379C}. For simplicity, we fix $\log_{10} U = -1$ in the models, roughly reflecting the best-fit value of the full \textsc{bagpipes} SED models (\cref{tab:UV_optical_properties}).

Given that stellar populations older than $10 \, \mathrm{Myr}$ become subdominant in terms of the ionising photon production, even with the inclusion of binary stars \citep[e.g.][]{2017PASA...34...58E}, we consider a simple model with a single burst of star formation aged either $1 \, \mathrm{Myr}$ or $10 \, \mathrm{Myr}$. Since the standard assumption is for the total nebular emission emerging from a model galaxy to be the superposition of emission from individual \HII regions associated with every SSP, scaled linearly by their ionising flux \citep{2001MNRAS.323..887C}, this simple model predicts a linear relation between SFR (which we average over a $10 \, \mathrm{Myr}$ window here) and emission-line luminosities, including $L_\text{\OIII}$ \citep[as also in][]{2020ApJ...903..150J}.

Assuming equal nebular and stellar metallicity, we find a monotonic increase of the $L_\text{\OIII} / \text{SFR}$ ratio with metallicity, in agreement with previous dedicated photoionisation modelling efforts \citep{2020ApJ...896...93H, 2022MNRAS.515.1751W}. However, we note that the quantitative predictions of our modelling approach are considerably different: in particular, for stellar ages of $1 \, \mathrm{Myr}$ our model is able to reach SFR-normalised \OIIIeightyeight luminosities of up to $\ssim 10^8 \, \mathrm{L_\odot}$ even at moderate metallicities of $Z = 0.2 \, \mathrm{Z_\odot}$. Importantly, this could help resolve the previously observed tension between observed $L_\text{\OIII} / \text{SFR}$ ratios and photoionisation models, where all but (near-)solar metallicity models struggled to explain the bright observed \OIIIeightyeight lines \citep{2022MNRAS.515.1751W}, even though in the most extreme case (COS-3018555981) direct-$T_e$ measurements with JWST later pointed towards a more modest $10$-$20\%$ oxygen abundances instead \citep{2025MNRAS.539.2463S}.

One potentially key difference with the approach adopted here is that these previous model approaches \citep[as well as the][ metallicity diagnostic]{2020ApJ...903..150J} relied on a single conversion factor to assign a SFR to a given model \HII region based on its (ionising) flux, typically based on the \Halpha line or UV continuum strength. In reality, however, these conversions depend sensitively on both IMF and metallicity \citep[e.g.][]{2012ARA&A..50..531K}, since more massive and/or metal-poor stars yield a higher rate of (ionising) UV photons, and hence \Halpha flux, per unit SFR \citep{2023ApJ...954..157S}. Here, instead, the mass-to-light ratio is self-consistently determined taking into account the IMF and metallicity of the stellar population.

Between the adopted $1 \, \mathrm{Myr}$ and $10 \, \mathrm{Myr}$ stellar ages and $5$-$50\%$ range in metallicity, the single-burst model considered here is able to reproduce most \OIIIeightyeight observations within the scatter of the \citet{2014A&A...568A..62D} relation for metal-poor dwarf galaxies. The observed \OIII luminosity and combined $\text{SFR}_{10}$ of \JGSzeleven falls just above the relation, closest to the $1 \, \mathrm{Myr}$ model curves with $5\%$ solar metallicity. The difference with our SED-based metallicity estimate can be explained by the more complex SFH adopted in our SED model, rather than a single burst of star formation (\cref{ssec:Discussion:SPS_modelling}).

One important caveat with the simple model presented for illustrative purposes here is that the \OIIIeightyeight emission may (partially) arise in a lower-density medium compared to the assumed $100 \, \mathrm{cm^{-3}}$. In such diffuse ionised gas, whose presence perhaps is indeed suggested by density-sensitive FIR line ratios \citep{2023MNRAS.521.2526K, 2024ApJ...977L...9Z, 2025ApJ...991L..38U, 2025ApJ...993..204H, 2025arXiv250702053M}, the $L_\text{\OIII} / \text{SFR}$ ratio is predicted to further increase \citep{2020ApJ...896...93H}. Further exploring this and other potential shortcomings of ad-hoc photoionisation models and their implementation in SED-fitting codes, however, is beyond the scope of this work. Direct forward modelling of observables based on detailed hydrodynamical zoom-in simulations coupled to chemical evolution models \citep[e.g.][]{2019MNRAS.487.5902K, 2022MNRAS.510.5603K, 2020MNRAS.496.5160L, 2020MNRAS.498.5541A, 2022MNRAS.513.5621P, 2023MNRAS.520L..16K, 2025A&A...704A..39K, 2023ApJ...953..140N, 2025arXiv250512397N} will provide a key avenue to approach this issue in the future.

\needspace{2.5cm}
\section{Conclusions}
\label{sec:Conclusions}

We have presented new ALMA observations of \OIIIeightyeight and dust-continuum emission in \JGSzeleven. Complementing recent ALMA detections of \OIII in the most luminous known galaxies beyond $z = 10$ \citep{2024ApJ...977L...9Z, 2025ApJ...988...19S, 2025A&A...696A..87C}, our results provide new insights into the gas, metal, and dust content of a somewhat fainter, more typical $z \approx 11$ galaxy. We summarise our main findings as follows:
\begin{enumerate}
    \item The nominal detection of \OIIIeightyeight at $\text{SNR} = 4.5$ ($\ssim 96.8\%$ purity) confirms the NIRSpec-derived redshift of \JGSzeleven based on the detection of multiple, faint UV and optical emission lines, which we refine to $z_\text{\OIII} = 11.1221 \pm 0.0006$. We find the line emission to be spectrally narrow ($\Delta v = 29 \pm 14 \, \mathrm{km \, s^{-1}}$), spatially extended, and potentially slightly offset in the direction of a fainter companion galaxy seen in the NIRCam imaging. We measure a luminosity of $L_\text{\OIII} = (1.1 \pm 0.3) \times 10^{8} \, \mathrm{L_\odot}$.
    \item Despite receiving one of the deepest cumulative ALMA integrations for a single $z > 10$ source so far ($19.2 \, \mathrm{h}$ on-source time in band 6 and 7), the dust-continuum emission remains undetected at the level of $S_\nu < 9.0 \, \mathrm{\upmu Jy}$ ($3\sigma$). This stringent non-detection allows us to place an upper limit on the IR luminosity of $L_\text{IR} < 3.1 \times 10^{10} \, \mathrm{L_\odot}$ (assuming $T_\text{dust} = 50 \, \mathrm{K}$) or an obscured SFR of $\text{SFR}_\text{IR} \lesssim 6 \, \mathrm{M_\odot \, yr^{-1}}$.
    \item Galvanised by an independently confirmed spectroscopic redshift, we performed joint SED modelling of the rich collection of spectrophotometric JWST data of \JGSzeleven, including deep measurements from NIRCam, NIRSpec and now also MIRI \citep{2025A&A...696A..57O}. As first pointed out by \citet{2024ApJ...976..160H}, this indicates that the neighbouring source is likely a low-mass companion ($M_* \approx 10^{7} \, \mathrm{M_\odot}$) which we find likely to exhibit strong nebular emission due to a recent star formation episode, whereas the main galaxy ($M_* \approx 2.8 \times 10^{8} \, \mathrm{M_\odot}$) prefers a more complex SFH resulting in combination of strong emission lines and a mild Balmer break.
    \item The accurate redshift measurement by ALMA confirms its value is reduced compared to initial photometric estimates by as much as $\Delta z \approx 0.5$, necessitating the presence of strong DLA absorption. The photometric bias can be partially circumvented with careful SED modelling, though spectroscopic constraints remain vital. Seemingly extreme in this compact object, the required neutral gas reservoir is in line with expectations based on the dynamical mass and dust attenuation, provided the absorbing gas has at most around $2\%$ solar metallicity. The implied gas surface density places \JGSzeleven in a highly star-bursting regime, although a dominant contribution of molecular gas could also still render it consistent with the KS relation.
    \item The bright \OIII line implies substantial metal enrichment has already taken place only $400 \, \mathrm{Myr}$ after the Big Bang, with an oxygen abundance likely at the level of approximately $20$-$30\%$ solar. Similar to GHZ2 \citep{2024ApJ...977L...9Z} and JADES-GS-z14-0 \citep{2025ApJ...988...19S, 2025A&A...696A..87C}, \JGSzeleven is consistent with the \citet{2014A&A...568A..62D} relation between \OIIIeightyeight luminosity and SFR calibrated on local, metal-poor dwarf galaxies.
\end{enumerate}

\noindent We conclude that ALMA is now providing a vital complementary view on the ISM conditions in the earliest galaxies discovered by JWST. Especially the \OIIIeightyeight line is starting to prove a bright and faithful tracer of recent SFR. Deeper integrations of \JGSzeleven in particular could provide efficient kinematical mapping of the two spatially resolved components seen in the rest-frame UV and provide new, highly relevant insights into the ionised-gas dynamics. Combined with other FIR transitions, ALMA will assist JWST over the coming years in unravelling the physical mechanisms shaping galaxy formation at Cosmic Dawn.

\section*{Acknowledgements}

We thank P.~Jakobsen for valuable discussions, and the MIRI European Consortium for publicly releasing the MIDIS imaging. This paper makes use of the following Atacama Large Millimeter/submillimeter Array (ALMA) data: ADS/JAO.ALMA\#2023.1.00336.S. ALMA is a partnership of the European Southern Observatory (ESO; representing its member states), NSF (USA) and NINS (Japan), together with NRC (Canada), MOST and the Academia Sinica Institute of Astronomy and Astrophysics (ASIAA; Taiwan), and KASI (Republic of Korea), in cooperation with the Republic of Chile. The Joint ALMA Observatory is operated by ESO, AUI/the National Radio Astronomical Observatory (NRAO) and the National Astronomical Observatory of Japan (NAOJ). This work is furthermore based on observations made with the National Aeronautics and Space Administration (NASA)/European Space Agency (ESA)/Canadian Space Agency (CSA) JWST. The data were obtained from the Mikulski Archive for Space Telescopes at the Space Telescope Science Institute (STScI), which is operated by the Association of Universities for Research in Astronomy, Inc., under NASA contract NAS 5-03127 for JWST. These observations are associated with programmes 1180, 1210, 1283, 1895, 1963, and 3215. The authors acknowledge the FRESCO and JEMS teams for developing their observing programme with a zero-exclusive-access period. JW gratefully acknowledges support from the Cosmic Dawn Center through the DAWN Fellowship. The Cosmic Dawn Center (DAWN) is funded by the Danish National Research Foundation under grant No. 140. RS acknowledges support from a Science and Technology Facilities Council (STFC) Ernest Rutherford Fellowship (ST/S004831/1). WMB gratefully acknowledges support from DARK via the DARK fellowship. This work was supported by a research grant (VIL54489) from VILLUM FONDEN. SA acknowledges grant PID2021-127718NB-I00 funded by the Spanish Ministry of Science and Innovation/State Agency of Research (MICIN/AEI/ 10.13039/501100011033). TJLCB acknowledges support from the Knut and Alice Wallenberg foundation through grant No. KAW 2020.0081. AJB, JC, and AS acknowledge funding from the ``FirstGalaxies'' Advanced Grant from the European Research Council (ERC) under the European Union’s Horizon 2020 research and innovation programme (Grant agreement No. 789056). SCa acknowledges support by European Union’s HE ERC Starting Grant No. 101040227 - WINGS. ECL acknowledges support of an STFC Webb Fellowship (ST/W001438/1). DJE is supported as a Simons Investigator. DJE, JMH, and BER acknowledge support from the NIRCam Science Team contract to the University of Arizona, NAS5-02015, and JWST programme 3215. Support for programme 3215 was provided by NASA through a grant from STScI, which is operated by the Association of Universities for Research in Astronomy, Inc., under NASA contract NAS 5-03127. This work has received funding from the Swiss State Secretariat for Education, Research and Innovation (SERI) under contract number MB22.00072. GCJ, RM, and JS acknowledge support by the STFC, by the ERC through Advanced Grant 695671 ``QUENCH'', and by the UK Research and Innovation (UKRI) Frontier Research grant RISEandFALL. RM also acknowledges funding from a research professorship from the Royal Society. PGP-G acknowledges support from grant PID2022-139567NB-I00 funded by Spanish Ministerio de Ciencia e Innovaci\'on MCIN/AEI/10.13039/501100011033, FEDER, UE. ST acknowledges support by the Royal Society Research Grant G125142. H\"U acknowledges funding by the European Union (ERC APEX, 101164796). DW is funded by the European Union (ERC, HEAVYMETAL, 101071865). Views and opinions expressed are however those of the authors only and do not necessarily reflect those of the European Union or the ERC Executive Agency. Neither the European Union nor the granting authority can be held responsible for them. This work has extensively used \textsc{casa} \citep{2007ASPC..376..127M}, developed by an international consortium of scientists based at NRAO, ESO, NAOJ, ASIAA, CSIRO Astronomy and Space Science (CSIRO/CASS), and the Netherlands Institute for Radio Astronomy (ASTRON), under the guidance of NRAO. This work has also relied on the following software packages: the \textsc{scipy} library \citep{Jones2001}, its packages \textsc{numpy} \citep{2011CSE....13b..22V} and \textsc{matplotlib} \citep{Hunter2007}, \textsc{astropy} \citep{2013A&A...558A..33A, 2018AJ....156..123A}, \textsc{bagpipes} \citep{2018MNRAS.480.4379C, 2019MNRAS.490..417C}, \textsc{cloudy} \citep{2017RMxAA..53..385F, 2023RMxAA..59..327C}, \textsc{emcee} \citep{2013PASP..125..306F}, \textsc{forcepho}, \textsc{(py)multinest} \citep{2009MNRAS.398.1601F, 2014A&A...564A.125B}, and \textsc{stpsf} \citep{2014SPIE.9143E..3XP}.

\clearpage
\bibliographystyle{apj}
\bibliography{ALMA_GS-z11.bib}

\begin{appendix}

\section{Photometric measurements}
\label{app:Photometric_measurements}
\begin{figure*}
	\centering
	\includegraphics[width=\linewidth]{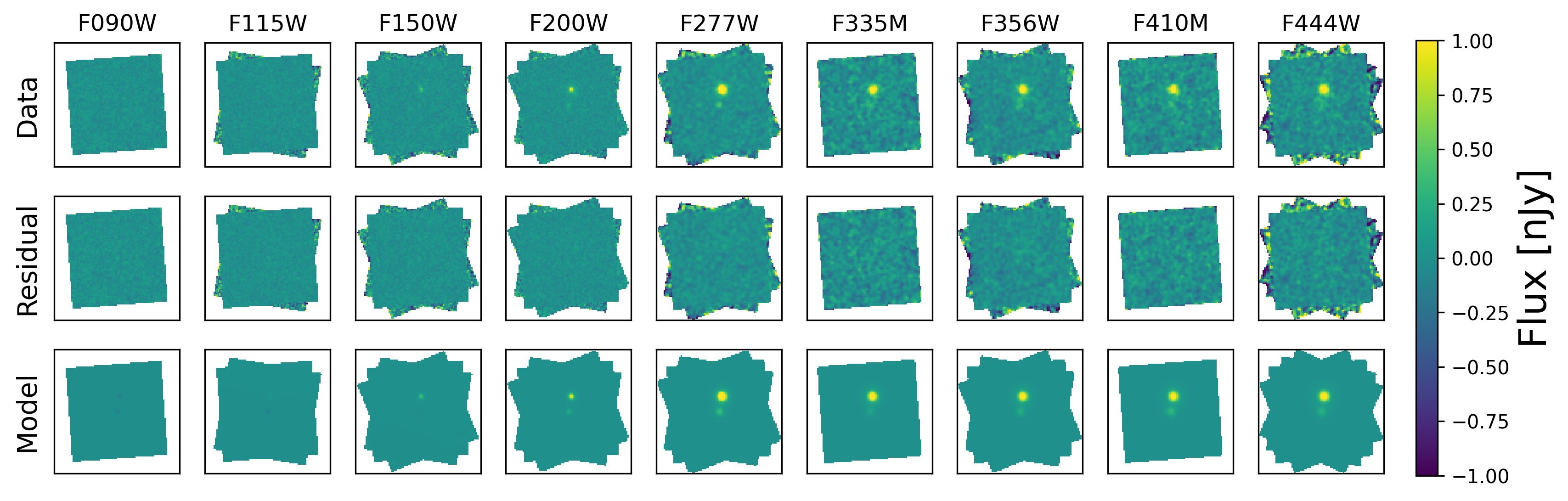}
	\caption{\textsc{forcepho} modelling of \JGSzeleven. \textit{Top row}: Cutouts of the observed NIRCam images (in detector coordinates, with slightly different orientation compared to \cref{fig:SEDs}) in a $\sim 1\arcsec \times 1\arcsec$ area around \JGSzeleven. Each column shows one of the $13$ available filters. \textit{Middle row}: Residuals between data and model. \textit{Bottom row}: PSF-convolved \textsc{forcepho} model images. While \textsc{forcepho} directly fits to individual exposures (\cref{app:Photometric_measurements}), the cutouts shown here are mosaiced for visualisation purposes.
	}
	\label{fig:ForcePho_scene}
\end{figure*}
\begin{figure*}
	\centering
	\includegraphics[width=\linewidth]{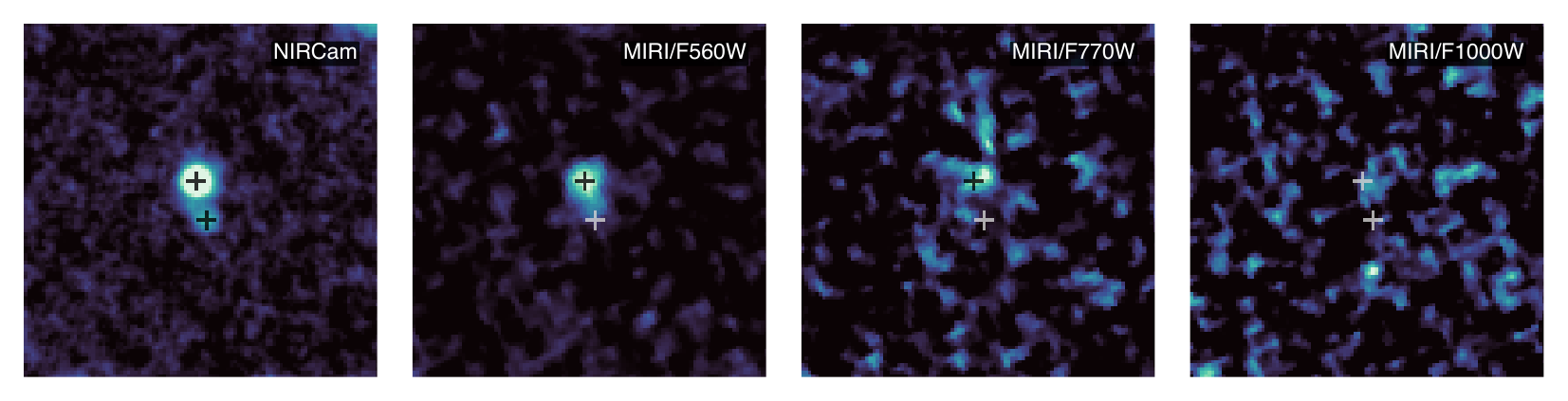}
	\caption{Comparison between NIRCam and MIRI imaging of \JGSzeleven. The NIRCam image (left) is an inverse-variance weighted stack of PSF-matched images from all bands with firm continuum detections, starting at F182M and going redwards. Black and white crosses show the NIRCam-based centroids of the two components (A and B) of \JGSzeleven.
	}
	\label{fig:MIRI_poststamps}
\end{figure*}

As in \citet{2025Natur.639..897W}, we obtained aperture photometry in small circular apertures, including in the available HST imaging (\cref{ssec:Observations:HST_JWST}) taken with the Advanced Camera for Surveys (ACS) and Wide Field Camera 3 (WFC3). Given the compact spatial size of \JGSzeleven, we considered `CIRC1' and `CIRC2' apertures, respectively $0.2\arcsec$ and $0.3\arcsec$ in diameter. In addition, we performed full-scene modelling of the NIRCam imaging in the direct surroundings of \JGSzeleven with \textsc{forcepho} \citetext{B.~D.~Johnson et al. in prep.}. \textsc{forcepho} models each source with a single intrinsic \citeauthor{1963BAAA....6...41S} profile, from which model images in all filters (each with freely varying normalisation) are obtained through convolution with their respective PSFs \citep[e.g.][]{2023NatAs...7..611R, 2023ApJ...952...74T, 2025NatAs...9..141B}. Crucially, we fit directly to all available individual NIRCam exposures, thereby avoiding correlated noise between adjacent pixels in drizzled mosaic images and enabling us to probe scales smaller than individual pixels.

The results are summarised in \cref{tab:Photometry} and shown in \cref{fig:ForcePho_scene}, from which it is apparent that both the main component and the fainter companion drop out of the F115W filter, consistent with $z \approx 11$. We measure sources A and B to have half-light radii of $29 \pm 1 \, \mathrm{mas}$ and $22 \pm 8 \, \mathrm{mas}$, respectively, translating to $114 \pm 8 \, \mathrm{pc}$ and $86 \pm 30 \, \mathrm{pc}$ at $z = 11.1$ (\cref{tab:Source_properties}). As expected from these small sizes, the \textsc{forcepho} photometry agrees well with the CIRC1 and CIRC2 aperture photometry.

As noted in \citet{2025Natur.639..897W}, however, the photometric uncertainty from the \textsc{forcepho} fitting does not directly capture any imperfections in the sky background subtraction as a possible source of systematic uncertainty. Instead, the impact of this effect is quantified empirically by placing a large number of randomly placed empty apertures, whose scatter is incorporated into the CIRC1 and CIRC2 aperture photometry uncertainty estimates \citep[see][ for a detailed discussion]{2023ApJS..269...16R}. The CIRC1 photometry of the fainter component B across $10$ HST and JWST filters up to and including F115W (the reddest $z \approx 11$ dropout filter) yield $\chi^2 = 12.2$ or $p = 0.2731$ under the null hypothesis of containing zero flux, which implies they are statistically fully consistent with non-detections.

As for the MIRI imaging from MIDIS \citep{2025A&A...696A..57O}, shown in \cref{fig:MIRI_poststamps}, we performed bespoke aperture photometry to minimise contamination, using radii of $0.27\arcsec$ in the F560W, $0.25\arcsec$ in F770W and F1000W filters for component A (listed as CIRC2 in \cref{tab:Photometry}) and $0.2\arcsec$ for all filters in component B (listed as CIRC1). The background was estimated in the nearby region, while the uncertainty was computed from non-adjacent pixels (separated by at least 5 pixels) to account for correlated noise, following the method described in \citet{2023ApJ...951L...1P}. These measurements were corrected to obtain total fluxes using aperture corrections derived with \textsc{stpsf} \citep[formerly WebbPSF;][]{2014SPIE.9143E..3XP}.
\begingroup
    \setlength{\tabcolsep}{8pt} 
    \renewcommand{\arraystretch}{1.25} 
    \begin{deluxetable}{cccccccc}
        \tabletypesize{\footnotesize}
        \tablecaption{Photometry of \JGSzeleven\label{tab:Photometry}}
        \startdata
        \tablehead{\colhead{Instrument} & \colhead{Filter} & \multicolumn{3}{c}{Component A} & \multicolumn{3}{c}{Component B}}
\\
& & \textsc{forcepho} & CIRC1 & CIRC2 & \textsc{forcepho} & CIRC1 & CIRC2
\\
\midrule
HST/ACS & F435W & \dots & $-0.42 \pm 0.61$ & $-1.02 \pm 0.90$ & \dots & $-0.02 \pm 0.61$ & $-0.80 \pm 0.90$
\\
& F606W & \dots & $0.09 \pm 0.47$ & $0.58 \pm 0.59$ & \dots & $-0.33 \pm 0.47$ & $0.17 \pm 0.59$
\\
& F775W & \dots & $0.63 \pm 0.51$ & $1.53 \pm 0.57$ & \dots & $-0.39 \pm 0.51$ & $-0.39 \pm 0.56$
\\
& F814W & \dots & $0.30 \pm 2.07$ & $-0.68 \pm 2.41$ & \dots & $-4.13 \pm 2.05$ & $-3.71 \pm 2.41$
\\
& F850LP & \dots & $0.42 \pm 1.25$ & $0.12 \pm 1.73$ & \dots & $0.45 \pm 1.23$ & $0.88 \pm 1.73$
\\
HST/WFC3 & F105W & \dots & $-2.58 \pm 0.76$ & $-3.10 \pm 0.75$ & \dots & $-1.07 \pm 0.76$ & $-1.87 \pm 0.74$
\\
& F125W & \dots & $-3.32 \pm 1.30$ & $-3.99 \pm 1.23$ & \dots & $-1.05 \pm 1.31$ & $-2.00 \pm 1.24$
\\
& F140W & \dots & $-1.45 \pm 1.13$ & $-1.96 \pm 1.50$ & \dots & $0.61 \pm 1.12$ & $0.08 \pm 1.50$
\\
& F160W & \dots & $6.01 \pm 1.38$ & $5.67 \pm 1.33$ & \dots & $-0.07 \pm 1.37$ & $-0.81 \pm 1.30$
\\
JWST/NIRCam & F090W & $-0.70 \pm 0.40$ & $0.05 \pm 0.68$ & $-0.08 \pm 0.87$ & $-0.48 \pm 0.46$ & $-0.85 \pm 0.69$ & $-1.67 \pm 0.87$
\\
& F115W & $0.26 \pm 0.31$ & $1.18 \pm 0.51$ & $1.64 \pm 0.75$ & $-1.08 \pm 0.36$ & $-0.78 \pm 0.50$ & $-0.95 \pm 0.73$
\\
& F150W & $5.99 \pm 0.36$ & $6.18 \pm 0.48$ & $7.49 \pm 0.67$ & $0.74 \pm 0.33$ & $0.83 \pm 0.47$ & $0.72 \pm 0.65$
\\
& F182M & $16.89 \pm 0.84$ & $16.26 \pm 0.97$ & $17.56 \pm 1.50$ & $2.20 \pm 0.60$ & $4.47 \pm 1.00$ & $4.85 \pm 1.52$
\\
& F200W & $16.19 \pm 0.81$ & $15.69 \pm 0.78$ & $18.04 \pm 0.90$ & $2.89 \pm 0.45$ & $4.15 \pm 0.53$ & $4.49 \pm 0.66$
\\
& F210M & $16.73 \pm 0.84$ & $16.92 \pm 1.17$ & $18.22 \pm 1.77$ & $0.86 \pm 0.60$ & $2.19 \pm 1.15$ & $1.83 \pm 1.75$
\\
& F277W & $17.97 \pm 0.90$ & $16.96 \pm 0.85$ & $19.38 \pm 0.97$ & $3.24 \pm 0.41$ & $3.13 \pm 0.38$ & $3.12 \pm 0.43$
\\
& F335M & $14.90 \pm 0.87$ & $14.18 \pm 0.75$ & $16.61 \pm 1.08$ & $2.41 \pm 0.83$ & $3.43 \pm 0.74$ & $3.91 \pm 1.06$
\\
& F356W & $16.51 \pm 0.83$ & $15.05 \pm 0.75$ & $17.10 \pm 0.85$ & $2.41 \pm 0.48$ & $3.16 \pm 0.37$ & $3.92 \pm 0.67$
\\
& F410M & $15.49 \pm 0.85$ & $14.01 \pm 0.86$ & $16.35 \pm 0.99$ & $2.83 \pm 0.80$ & $4.49 \pm 0.86$ & $4.20 \pm 1.00$
\\
& F430M & $15.70 \pm 2.99$ & $15.27 \pm 2.08$ & $15.88 \pm 3.24$ & $2.59 \pm 2.57$ & $7.52 \pm 2.10$ & $8.47 \pm 3.27$
\\
& F444W & $18.09 \pm 0.90$ & $16.42 \pm 0.82$ & $18.63 \pm 0.93$ & $2.98 \pm 0.73$ & $3.67 \pm 0.64$ & $5.30 \pm 0.79$
\\
& F460M & $20.01 \pm 3.71$ & $18.33 \pm 4.42$ & $19.96 \pm 4.08$ & $2.17 \pm 4.16$ & $4.79 \pm 4.42$ & $4.29 \pm 4.17$
\\
& F480M & $18.98 \pm 3.31$ & $18.64 \pm 2.87$ & $21.15 \pm 3.40$ & $0.00 \pm 2.46$ & $2.14 \pm 2.77$ & $0.66 \pm 3.40$
\\
JWST/MIRI & F560W & \dots & \dots & $28.39 \pm 1.85$ & \dots & $10.17 \pm 1.13$ & \dots
\\
& F770W & \dots & \dots & $18.80 \pm 2.35$ & \dots & $5.26 \pm 3.01$ & \dots
\\
& F1000W & \dots & \dots & $13.92 \pm 6.54$ & \dots & $3.07 \pm 6.39$ & \dots
        \enddata
        \tablecomments{
            Flux densities $F_\nu$ in nJy are listed for the main component (A) and companion (B) in each filter.
        }
    \end{deluxetable}
\endgroup
\FloatBarrier
\clearpage

\section{NIRSpec/PRISM covariance matrix}
\label{app:Covariance_matrix}

To obtain our final, combined NIRSpec/PRISM spectrum from the $186$ individual sub-exposures (\cref{ssec:Observations:HST_JWST}), we used the combination and filtering process as described in \citet{2024ApJ...976..160H} and \citet{2025Natur.639..897W}. Following these works, we also constructed a covariance matrix $\mathbf{C}$ using a bootstrapping procedure with $5000$ iterations. In \cref{fig:Covariance_matrix_diagonal}, we show the standard deviation on the flux density $F_\lambda$ in the NIRSpec/PRISM spectrum, defined as the square root of the diagonal elements of the covariance matrix, $\sigma_i^2 = C_{ii}$. Also shown is the behaviour of the correlation matrix, $\rho_{ij} = C_{ij}/(\sigma_i \sigma_j)$, for entries with a given offset $x \equiv |i-j|$ from the diagonal. Directly adjacent wavelength bins clearly have strongly correlated noise, though this quickly drops towards offsets of more than $2$ wavelength bins.
\begin{figure}
	\centering
	\includegraphics[width=0.95\linewidth]{"Covariance_matrix_diagonal"}
	\caption{Standard deviation on the flux density $F_\lambda$ in the NIRSpec/PRISM spectrum, where $\sigma_i$ is based on the diagonal elements of the covariance matrix $\sigma_i^2 = \sqrt{ C_{ii} }$. The inset show the distribution of off-diagonal elements in the correlation matrix, $\rho_{ij} = C_{ij}/(\sigma_i \sigma_j)$, for entries offset by $i - j$ bins from the diagonal ($i - j = 0$ being the central diagonal). The solid green line represents the median and the shading ranges between the \nth{16} and \nth{84} percentiles. In both panels, a solid black line shows the model covariance (see also \cref{fig:Covariance_matrix}).
	}
	\label{fig:Covariance_matrix_diagonal}
\end{figure}

\begin{figure*}
	\centering
	\includegraphics[width=0.95\linewidth]{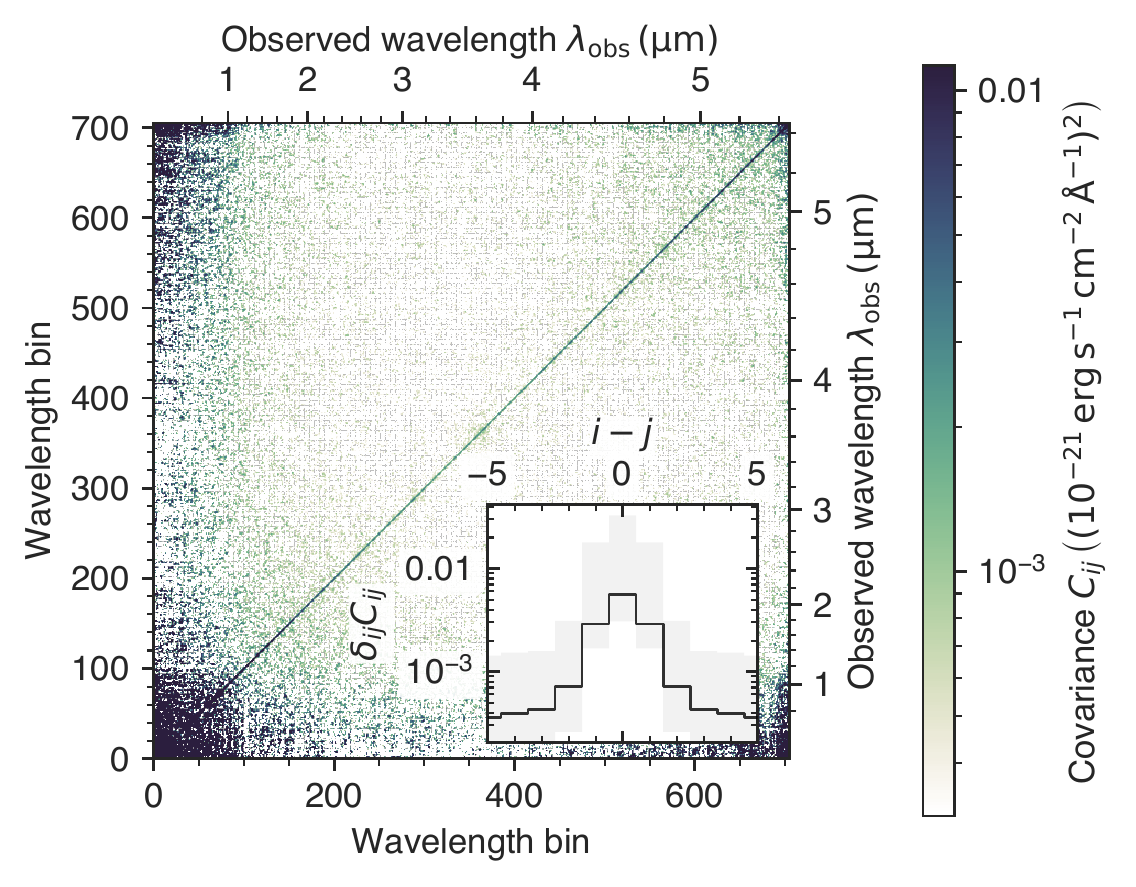}
	\caption{\textit{Left}: Bootstrapped covariance matrix $\mathbf{C}$ of the NIRSpec/PRISM spectrum. \textit{Centre}: model covariance (see text for details). \textit{Right}: absolute value of the residuals.
	}
	\label{fig:Covariance_matrix}
\end{figure*}

To prevent issues with the inversion of the covariance matrix due to noise artefacts, we model the covariance in the $1.3$-$5.3 \, \mathrm{\upmu m}$ region through a combination of two power laws for the diagonal entries (as shown in \cref{fig:Covariance_matrix_diagonal}), which are modulated away from the diagonal by one narrow squared exponential term, $\propto \exp(-x^2 / (2b_1^2))$, and one broad exponential, $\propto \exp(-x / b_2))$ (as shown in the inset in \cref{fig:Covariance_matrix_diagonal}). The full covariance matrix $\mathbf{C}$, alongside the model and residuals, is shown in \cref{fig:Covariance_matrix}.

\section{Stellar population synthesis modelling}
\label{app:SPS_modelling}
\begin{figure*}[!b]
	\centering
	\includegraphics[width=\linewidth]{"Corner_ultra-deep-gs-3215_20130158_v5.1_extr3_lines_phot_ForcePho_v1.0d.0_bursty-continuity-DM_bpass_v2.3.a+02_CF00_FIR"}
	\caption{Posterior distributions of the main physical parameters (\cref{tab:UV_optical_properties}), obtained from \textsc{bagpipes} fits to the spectrum and photometry of \JGSzeleven (component A).
	}
	\label{fig:Corner plot}
\end{figure*}

In this appendix, we describe our SED fitting procedure with \textsc{bagpipes} v1.3.2 \citep{2018MNRAS.480.4379C, 2019MNRAS.490..417C}, which largely follows the modelling approach of \citet{2025MNRAS.536...27W, 2025Natur.639..897W}. We considered stellar models including binary stars from the v2.3 Binary Population and Spectral Synthesis \citep[\textsc{bpass};][]{2022MNRAS.512.5329B} library under their default \citet{2001MNRAS.322..231K} IMF with a stellar mass upper limit of $300 \, \mathrm{M_\odot}$, which are identical to the v2.2.1 models \citep{2017PASA...34...58E} except that the abundances of $\alpha$ elements are enhanced relative to iron, for discrete values of $[ \mathrm{\alpha/Fe} ] \in \left\{ -0.2, 0, 0.2, 0.4, 0.6 \right\} \, \mathrm{dex}$.

We revised the default implementation for DLA absorption in \textsc{bagpipes} to match the prescription within the \textsc{lymana\_absorption} code\footnote{Code available at \url{https://github.com/joriswitstok/lymana_absorption}.}, which is described in detail in \citet{2025MNRAS.536...27W}. Briefly, we model the absorption cross section of neutral hydrogen as the Voigt profile approximation from \citet{2006ApJ...645..792T}. Since the higher-order quantum-mechanical correction from \citet{2015MNRAS.446..264B} is only valid redwards of $1100 \, \Angstrom$, we apply the linear correction provided by \citet{2013ApJ...772..123L}. While the redshift of the foreground DLA system can be freely varied \citep[e.g.][]{2024A&A...690A..70T}, we opt to fix this to the systemic redshift throughout. When including DLA absorption, we assumed a log-uniform prior across a range of \HI column densities of $10^{19} \, \mathrm{cm^{-2}} < N_\text{\HI} < 10^{24} \, \mathrm{cm^{-2}}$.

As in \citet{2023MNRAS.522.6236T}, the SFH is non-parametric with $6$ bins in lookback time $t$. The first two bins are placed to cover $0 < t < 3 \, \mathrm{Myr}$ and $3 < t < 10 \, \mathrm{Myr}$ to be able to capture extreme line emission, whereas the remaining bin edges are spaced logarithmically out to $z = 20$. We adopted a `bursty-continuity' prior \citep{2019ApJ...876....3L}, where the logarithmic ratio of SFR in adjacent bins is modelled as a Student's-$t$ distribution with $\nu = 2$ degrees of freedom and scale $\sigma = 1.0$, though we shifted its mean away from zero to follow a physically motivated rising SFH based on dark matter halo accretion, as proposed by \citet[; see also \citealt{2025A&A...696A..87C}]{2025MNRAS.537.1826T}. As discussed in the main text, however, following \citet{2025ApJ...992..212W} we adopt a $\text{SFR}(z) \propto (1+z)^{-4.5}$ scaling based on halo mass abundance matching in the \textsc{abacus} $N$-body simulations \citep{2021MNRAS.508.4017M, 2024Natur.633..318C} instead of the \citet{2013MNRAS.435..999D} analytic approximation of dark matter halo accretion rates, which focussed on a lower-redshift regime than considered here.

The total stellar mass formed was varied across $0 < \log_{10} \left( M_* \, (\mathrm{M_\odot}) \right) < 15$ and stellar metallicity over a range of $0.0005 \, \mathrm{Z_\odot} < Z_* < 1.5 \, \mathrm{Z_\odot}$, both with log-uniform priors. Nebular emission is derived using the \citet{2017ApJ...840...44B} prescription, under the assumption that the gas-phase metallicity is equal to the stellar one, and calculated with \textsc{cloudy} v23.01 models \citep{2017RMxAA..53..385F, 2023RMxAA..59..327C} that are self-consistently irradiated by the appropriate stellar population, with the ionisation parameter as a free parameter ($-3 < \log_{10} U < -0.5$).

In light of recent solar abundance measurements \citep[e.g.][]{2022A&A...661A.140M}, as well as to remain consistent with the adopted solar metallicity of $Z_\odot = 0.02$ in \textsc{bpass}, we maintained the default \textsc{cloudy} solar abundance pattern from \citet{1989GeCoA..53..197A}, while adjusting the gas-phase iron abundance according to the $\alpha$ enhancement $[ \mathrm{\alpha/Fe} ]$ of each of the \textsc{bpass} v2.3 model sets. We also implemented the possibility of directly fitting emission line fluxes in \textsc{bagpipes}. Dust attenuation is included via a flexible \citet{2000ApJ...539..718C} law \citep[see][]{2019MNRAS.483.2621C}. Dust-continuum emission is incorporated according to the \citet{2007ApJ...657..810D} model with $q_\text{PAH} = 0.75\%$, a log-uniform prior on $U_\text{min}$ ($0.25 \leq U_\text{min} \leq 15$), and a uniform prior on $\gamma$ ($0 \leq \gamma \leq 0.4$).

In the case of component A of \JGSzeleven, the spectral resolution curve of the PRISM was calculated from detailed forward modelling of the NIRSpec instrument, taking into account the morphology and intra-shutter position of this component \citep{2024A&A...684A..87D}. We included a first-order Chebyshev polynomial correction to the spectroscopic data \citep{2019MNRAS.490..417C} to account for any minimal discrepancies between NIRSpec and NIRCam.

Under the assumption the observed data are normally distributed around the model, the spectral model log-likelihood $\ell_\text{spec}$ or the related goodness-of-fit statistic $\chi_\text{spec}^2$ are derived via the inverse covariance matrix $\mathbf{C}^{-1}$ (Appendix~\ref{app:Covariance_matrix}),
\begin{equation}
    \ell_\text{spec} = K_\text{spec} - \frac{1}{2} \mathbf{R}^T \mathbf{C}^{-1} \mathbf{R} = K_\text{spec} - \frac{1}{2} \chi^2 \, ,
\end{equation}
where $\mathbf{R}$ is the vectorised difference between observed and modelled flux density in the $i^\text{th}$ wavelength bin, respectively $F_{\lambda, i}^\text{obs}$ and $F_{\lambda, i}^\text{model}$,
\begin{equation}
    R_i = F_{\lambda, i}^\text{obs} - F_{\lambda, i}^\text{model} \, .
\end{equation}
For computational reasons, the constant $K_\text{spec}$ is derived from the determinant of the covariance matrix $\mathbf{C}$ via its $N$ eigenvalues $\lambda_i$:
\begin{align}
    K_\text{spec} & = - \frac{1}{2} \ln \left( 2 \pi \det(\mathbf{C}) \right) = - \frac{1}{2} \ln \left( 2 \pi \, \Pi_{i=0}^{N} \lambda_i \right) \nonumber
    \\
    & = - \frac{1}{2} \left( \ln (2 \pi) + \Sigma_{i=0}^{N} \ln (\lambda_i) \right) \, .
\end{align}
To ensure convergence of the \textsc{(py)multinest} sampler \citep{2009MNRAS.398.1601F, 2014A&A...564A.125B}, we masked the noise-dominated wavelength regions $\lambda_\text{obs} < 1.3 \, \mathrm{\upmu m}$ and $\lambda_\text{obs} > 5.3 \, \mathrm{\upmu m}$ (cf. \cref{fig:Covariance_matrix_diagonal}). The main results of the SED fitting are tabulated in \cref{tab:UV_optical_properties}. For component A, the full posterior distributions of the main parameters of interest are shown in \cref{fig:Corner plot}.

\end{appendix}
\end{document}